%% file: main.tex
\documentclass[conference,a4paper]{IEEEtran}

\input{preamble/packages}
\input{preamble/commands}

\input{preamble/acronyms}

\ifCLASSINFOpdf
\else
\fi
\newboolean{full}
\setboolean{full}{true} 

\newboolean{mmdec}
\setboolean{mmdec}{true} 

\newboolean{mmdec_gen}
\setboolean{mmdec_gen}{true} 

\newboolean{mmrelay}
\setboolean{mmrelay}{true} 

\newboolean{comib}
\setboolean{comib}{false} 

\newboolean{ibcsi}
\setboolean{ibcsi}{false} 

\begin{document}
%
\title{On Mismatched Oblivious Relaying}
%
%
%

\author{
	\IEEEauthorblockN{Michael Dikshtein, Nir Weinberger,  and Shlomo Shamai (Shitz)} \IEEEauthorblockA{Department of Electrical and Computer Engineering, Technion, Haifa 3200003, Israel}
	\IEEEauthorblockA{ 		
		Email: \{michaeldic@, nirwein@, sshlomo@ee.\}technion.ac.il
}
}
\maketitle

\begin{abstract}
\input{sections/abstract}
\end{abstract}


%
\IEEEpeerreviewmaketitle

\section{Introduction}
\input{sections/introduction}

\ifthenelse{\boolean{mmdec}}{
\section{Information Bottleneck Channel with Mismatched Decoder }
\label{section:mmdec}
\input{./sections/mmdec}

}
{}

\ifthenelse{\boolean{mmdec_gen}}{
\subsection{Continuous-Alphabet Memoryless Channels }
\input{./sections/mmdec_gen}
}
{}

\ifthenelse{\boolean{mmrelay}}{
\section{Information Bottleneck Channel with Mismatched Relay }
\label{section:mmrelay}
\input{./sections/mmrib_main_results}

}{}

\ifthenelse{\boolean{comib}}{
\section{Compound Information Bottleneck Channel}
\input{./sections/compound}

}{}

\ifthenelse{\boolean{ibcsi}}{
\section{Information Bottleneck Channel with Side Information}
\input{./sections/ibcsi}
}{}



\section{Summary and Outlook}
\input{./sections/conclusions}
\label{section:conclusion}


\section*{Acknowledgment}
The work has been supported by the 
	European Union's Horizon 2020 Research And Innovation Programme,
	grant agreement no. 694630, by the ISF under Grant 1791/17, and by the WIN consortium via the
	Israel minister of economy and science.


%

\newpage

\bibliographystyle{ieeetr}
\bibliography{mmibbib}

\ifthenelse{\boolean{full}}
{
\appendices
\section{Notations Conventions}
\input{sections/notations}


\ifthenelse{\boolean{mmdec}}{
\section{Proof of achievability for \Cref{theorem:mmib_mismatched_decoder_rate}}
\label{appendix:theorem_mmib_mismatched_decoder_rate_direct_proof}
\input{./sections/mmib/lower_bound}

\section{Proof of ensemble tightness for \Cref{theorem:mmib_mismatched_decoder_rate}}
\label{appendix:theorem_mmib_mismatched_decoder_rate_converse_proof}
\input{./sections/mmib/converse}

\section{Proof of achievability for \Cref{theorem:mmib_mismatched_decoder_gmi_rate}}
\label{appendix:theorem_mmib_mismatched_decoder_gmi_rate_direct_proof}
\input{./sections/mmib/gmi_lower_bound_proof}
}{}


\ifthenelse{\boolean{mmrelay}}{
\section{Proof of achievability for \Cref{theorem:mmib_mismatched_relay_rate}}
\label{appendix:theorem_mmib_mismatched_relay_rate_direct_proof}
\input{./sections/mmrib/lower_bound}
}{}










}
{}



\ifCLASSOPTIONcaptionsoff
  \newpage
\fi

\end{document}

%% file: preamble/packages.tex
\usepackage{amsfonts}
\usepackage{amsmath}
\usepackage{amssymb}

\usepackage[ruled,vlined]{algorithm2e}

\usepackage{caption}
\usepackage{subcaption}

\usepackage{color,soul}

\usepackage[acronym]{glossaries}

\usepackage{cleveref} 

\usepackage{mathtools}

\usepackage{tikz}
\usetikzlibrary{shapes.multipart}

\usepackage{pgfplots}

%% file: preamble/commands.tex
\newcommand{\bv}[1]{\mathbf{#1}}
\newcommand{\epmf}[1]{\hat{\mathsf{P}}_{#1}}
\newcommand{\eqann}[2]{\overset{\mathclap{(\text{#2})}}{#1}} 
\newcommand{\eqannref}[1]{$(\text{#1})$}
\newcommand{\event}[1]{\mathcal{#1}}
\newcommand{\Exp}[2][]{\mathbb{E}_{#1} \left[#2\right]}
\newcommand{\field}[1]{\mathbb{#1}}
\newcommand{\indicator}[1]{\boldsymbol{1} \left\{#1 \right\}}
\newcommand{\kld}[2]{\mathsf{D} \left( #1 \| #2 \right)}
\newcommand{\minus}{\scalebox{0.75}[1.0]{\( - \)}}
\newcommand{\rv}[1]{\mathsf{#1}}
\newcommand{\pmf}[1]{\mathsf{P}_{#1}}
\newcommand{\pmfT}[1]{\tilde{\mathsf{P}}_{#1}}
\newcommand{\pmfQ}[1]{\mathsf{Q}_{#1}}
\newcommand{\Prob}[1]{\mathsf{P} \left\{#1 \right\}}
\newcommand{\set}[1]{\mathcal{#1}}
\newcommand{\typeclass}[2]{\set{T}^{(#1)} \left(#2 \right)}
\newcommand{\typset}[2]{\set{T}_{#1}^{(#2)}}

\newtheorem{theorem}{Theorem}
\newtheorem{lemma}[theorem]{Lemma}
\newtheorem{remark}{Remark}
\newtheorem{definition}{Definition}[section]

\DeclareMathOperator*{\argmax}{arg\,max}
\DeclareMathOperator*{\argmin}{arg\,min}

%% file: preamble/acronyms.tex
\newacronym{awgn}{AWGN}{additive white Gaussian noise}

\newacronym{bpsk}{BPSK}{binary phase shift keying}
\newacronym{cscg}{CSCG}{\textit{circularly symmetric complex Gaussian}}
\newacronym{csi}{CSI}{Channel State Information}
\newacronym{dm}{DM}{\textit{discrete memoryless}}
\newacronym{dmc}{DMC}{\textit{discrete memoryless channel}}
\newacronym{dms}{DMS}{\textit{discrete memoryless source}}
\newacronym{dmibc}{DM-IBC}{\textit{discrete memoryless information bottleneck channel}}
\newacronym{dmmmibc}{DM-MMIBC}{\textit{discrete memoryless mismatched information bottleneck channel}}
\newacronym{dmmmribc}{DM-MMRIBC}{\textit{discrete memoryless mismatched relay information bottleneck channel}}
\newacronym{dmmmrdibc}{DM-MMRDIBC}{\textit{discrete memoryless mismatched relay and decoder information bottleneck channel}}

\newacronym{gmi}{GMI}{\textit{generalized mutual information}}

\newacronym{ib}{IB}{\textit{information bottleneck}}
\newacronym{ibam}{IBAM}{\textit{information bottleneck alternating minimization}}

\newacronym{iff}{iff}{if and only if}
\newacronym{iid}{i.i.d.}{independent, identically distributed}

\newacronym{kl}{KL}{\textit{Kullback-Leibler}}
\newacronym{lln}{LLN}{law of large numbers}
\newacronym{mmib}{MMIB}{Mismatched Information Bottleneck}
\newacronym{pmf}{\textbf{pmf}}{\textit{probability mass function}}
\newacronym{rhs}{RHS}{right-hand side}

\newacronym{wlog}{w.l.o.g.}{without loss of generality}

%% file: sections/abstract.tex
We consider the problem of reliable communication over a \acrfull{dmc} with the help of a relay,  termed the \acrfull{ib} channel. There is no direct link between the source and the destination, and the information flows in two hops. The first hop is a noisy channel from the source to the relay. The second hop is a noiseless but limited-capacity backhaul link  from the relay to the decoder. We further assume that the relay is oblivious to the transmission codebook. We examine two mismatch scenarios. In the first setting, we assume the decoder is restricted to use some fixed decoding rule, which is mismatched to the actual channel. In the second setting, we assume that the relay is restricted to use some fixed compression metric, which is again mismatched to the statistics of the relay input. We establish bounds on the \textit{random-coding} capacity of both settings, some of which are shown to be ensemble tight.


%% file: sections/introduction.tex
In this paper, we consider the point-to-point oblivious relay channel \cite{Sanderovich2008a}, in which a relay observes the transmitted signal over a noisy channel and transmits digital information to the decoder via a limited-capacity link. Our primary focus in this paper is  achievable rate results under various \textit{mismatch} conditions, specifically using random codes. We consider both  mismatched decoding at the receiver as well as  mismatched compression of the relay. The motivation for this analysis is that network architectures with oblivious processing at the relays serve as the fundamental building blocks of modern communication systems. A recent and comprehensive summary on oblivious communication networks can be found in  \cite{EstellaAguerri2019}. The motivation for analyzing the mismatched case is that in many practical settings, .e.g., in the up-link of cellular communication with oblivious relays, the relay or the decoder only possess partial information regarding the statistical model of the channel or is restricted to operating with some specific decoding metric (and typically use it in a more elaborate channel decoding algorithm, such as  belief-propagation \cite{Richardson2008}).

The capacity of the oblivious relay channel is tightly connected to the \acrfull{ib}  problem \cite{Tishby1999}, which has been the subject of a recent extensive study, mainly due to its relation to current advances in machine learning, see e.g., \cite{zaidi2020information,Goldfeld2020}, and statistical learning \cite{Aguerri2021}.
In the classical rate-distortion theory of lossy source coding, a fidelity measure must be chosen that quantifies the quality of compression \cite{shannon1959coding}. 
The \acrshort{ib} method determines this distortion measure via an additional dependent random variable that captures the meaningful information in the data to be compressed, which can be thought of as contextual labeling of the data. Specifically, the compression quality under the \acrshort{ib} approach is assessed via the mutual information between the compressed representation and the additional variable. As it turns out, the resulting \acrshort{ib} matches precisely the capacity of the oblivious relay channel under consideration. In effect, it is also the single-letter rate-distortion formula for remote-source coding setting \cite{Dobrushin1962,Wolf1970} when the distortion measure is the log-loss \cite{Courtade2014b}. It was shown in \cite{Painsky2018} that minimizing the log-loss minimizes an upper bound to any choice of loss functions for binary classification problems. 
It can also be shown that log-loss actually bounds  general distortion measures (as stated by Linder, but to the best of our knowledge, it has not been published).


As is well known, the problem of mismatched decoding is notoriously challenging and is not fully resolved, even for standard point-to-point channels \cite{Scarlett2020}. Nonetheless, analysis of random codes under mismatched encoding or decoding leads to tractable achievable bounds, and so we adopt this analysis for the oblivious relay channel (or the \acrshort{ib} problem) studied in this paper.
Coding over a \acrshort{dmc} with mismatched decoder under the random coding regime was introduced independently in \cite{hui1983fundamental} and \cite{Csiszar1981}, where a lower bound (termed the \textit{LM rate}) on the capacity was derived. In \cite{Csiszar1995} it was shown that the LM bound is not tight. A more analytically tractable lower bound, termed \acrfull{gmi}, was proposed in \cite{kaplan1993information}, and the random coding ensemble tightness for the LM scheme was established in \cite{Merhav1994}. In our setting, the analysis of random codes is further motivated by the desire to model codes that are not adapted to a specific communication setting, by the obliviousness nature of the relay, and by security aspects typically involved in relay communication systems.

The outline of the rest of the paper and our contributions are as follows. 
In Sec. \ref{section:mmdec}, we consider the \acrshort{ib} problem with a mismatched decoder at the receiver and derive random coding (achievable) rates, both in the form of an LM bound, as well as a GMI bound. We then propose an algorithm for the computation of the achievable rate and exemplify its operation on a quaternary channel. We then extend the GMI rate to  continuous alphabet channels and demonstrate this result for a Gaussian fading channel. Afterward, in Sec. \ref{section:mmrelay}, we consider the setting of a relay with a mismatched compression rule. In Sec. \ref{section:conclusion}, we conclude the paper.

\paragraph*{Related Work}
A comprehensive summary on information-theoretic foundations of mismatched decoding and encoding is provided in \cite{Scarlett2020}.
Beyond channel coding, mismatch has also been studied in the context of source coding \cite{lapidoth1997,zhou2019}. A successive refinement setting constrained to Gaussian codebooks with minimal Euclidean distance encoding has been proposed in \cite{bai2022achievable}. An extension to general alphabets has been recently presented in \cite{Kanabar2022}.  
The global channel knowledge at the destination setting has been studied in the context of quantized distributed reception in \cite{Choi2015}, where a simple, complex binary sign quantization has been assumed. A distributive decoding communication network with \acrshort{bpsk} transmission over the \acrshort{awgn} channel has been considered in \cite{Brown2013}. Uniform quantization for OFDMA-based CRAN has been considered in \cite{liang2014joint}. Outage probability in the problem of distributed reception with hard decision exchanges has been considered in \cite{Wang2013}.

%% file: sections/mmdec.tex
\subsection{Discrete Memoryless Channels}
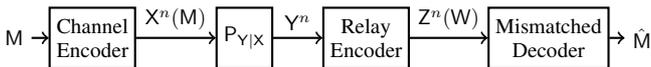
\begin{figure}[b]
    \centering
    \input{./figures/mmib/block_diagram}
    \caption{Oblivious Communication System with Mismatched Decoder}
    \label{fig:mmib_communication_diagram}
\end{figure}
In this section, we consider the 3-node point-to-point communication system with a relay depicted in \Cref{fig:mmib_communication_diagram}, in which the sender wishes to communicate a message $\rv{M}$ to the receiver with the help of the relay. We term this setting the \acrfull{dmibc} with mismatched decoder $\langle \set{X}, \pmf{\rv{Y}|\rv{X}},  \set{Y}, \set{Z},V(z|x) \rangle$. It consists of three finite sets $\set{X}$, $\set{Y}$, $\set{Z}$,  a collection of conditional \acrshort{pmf}s $\pmf{\rv{Y}|\rv{X}}$ on $\set{Y}$ (one for each input symbol $x$), and a decoding metric $V(z|x)$  on $\set{Z}$.

A $(2^{nR},2^{nB},n)$ code for the \acrshort{dmibc} with mismatched decoder consists of:
\begin{itemize}
    \item a message set $\set{M} = [1:2^{nR}]$,
    \item a representation set $\set{W} = [1:2^{nB}]$,
    \item an encoder that assigns a codeword $x^n(m)$ to each message $m \in \set{M}$,
    \item a relay source encoder that assigns an index $w \in \set{W}$ to each received sequence $y^n \in \set{Y}^n$, with the respective reconstructed sequence $z^n(w)$, and
    \item a decoder that assigns an estimate $\hat{m}$ or an error message e to each received representation index $w \in \set{W}$ according to some fixed mismatched metric 
    \begin{multline} \label{eq:mmdibc_decoding_rule}
        \hat{m}(z^n) = \argmax_{m \in \set{M}} V(z^n|x^n(m)) \\
        = \argmax_{m \in \set{M}} \prod_{i=1}^n V(z_i|x_i(m)).
    \end{multline}
\end{itemize}
It is assumed that the message $\rv{M}$ is uniformly distributed over the message set $\set{M}$.

\begin{definition}
    Consider a codebook of $2^{nR}$ $n$-dimensional sequences, $\{x^n(m)\}_{m=1}^{2^{nR}}$, where each sequence is generated at random with a memoryless \acrshort{pmf} $\pmf{\rv{X}}$ and independently of all other vectors. Also consider a compression codebook of $2^{nB}$ $n$-dimensional sequences, $\{z^n(w)\}_{w=1}^{2^{nB}}$, where each sequence is generated at random with a memoryless \acrshort{pmf} $\pmf{\rv{Z}}$ and independently of all other vectors. A pair of such channel-compression codebooks is termed a \textit{random codebook}. Let $P_e^{(n)} = \mathsf{P} \{\hat{\rv{M} } \neq \rv{M} \}$ denote the error probability averaged over the random codebooks.
     A rate $R$ is said to be \textit{achievable} for the \acrshort{dmibc} at  compression rate $B$, with mismatched decoding metric $V$, if 
     $\lim_{n\rightarrow \infty} P_e^{(n)} = 0$ under the mismatched decoding rule. The \textit{random coding capacity} $C_V(B)$ of the \acrshort{dmibc} with mismatched decoder is the supremum of all achievable rates of random codebooks. 
\end{definition}

Our main result of this section is stated in the following theorem, and it describes the respective LM rate \cite{hui1983fundamental,Csiszar1981} for the oblivious relay setting. 
\begin{theorem} \label{theorem:mmib_mismatched_decoder_rate}
    The random coding capacity of the \acrshort{dmibc} with mismatched decoder is
    \begin{equation} \label{eq:mmibdec_Jxz_outer_optimization_problem}
        \begin{aligned}
            &C_V(B) =
            && \max_{\pmf{\rv{X}}}\max_{\pmf{\rv{Z}|\rv{Y}}}
            && J_{V}(\rv{X};\rv{Z}) \\
            &&& \text{subject to}
            && I(\rv{Y};\rv{Z}) \leq B,
        \end{aligned}
    \end{equation}
    where
    \begin{equation} \label{eq:mmibdec_Jxz_optimization_problem}
        \begin{aligned}
            &J_V(\rv{X};\rv{Z}) = 
            &&\min_{\pmfQ{\rv{Z}|\rv{X}} \in \set{P}_{\set{X}\times \set{Z}}}
            && I( \pmf{\rv{X}}, \pmfQ{\rv{Z}|\rv{X}}) \\
            &&& \text{subject to} 
            && \sum_{x\in \set{X}} \pmfQ{}(x,z) = \pmf{\rv{Z}} (z), \\
            &&&&& \mkern-25mu \sum_{(x,z) \in \set{X} \times \set{Z}} \mkern-20mu  \pmfQ{} (x,z)  \log V(z|x) \geq -D,
        \end{aligned}
        \vspace{-5pt}
    \end{equation}
    with $\pmfQ{}(x,z) = \pmfQ{\rv{Z}|\rv{X}}(z|x) \cdot \pmf{\rv{X}}(x)  $ and
    \begin{equation} \label{eq:mmibdec_lm_rate_contraintD}
        D \triangleq - \sum_{x \in \set{X}}  \sum_{z \in \set{Z}}  \pmf{\rv{X} \rv{Z}}(x,z)  \log  V(z|x).
    \end{equation}
\end{theorem}

Theorem \ref{theorem:mmib_mismatched_decoder_rate} provides the \textit{exact} random coding capacity under mismatched decoding. In specific, both an upper bound and ensemble tightness are proved. 

\ifthenelse{\boolean{full}}
{
\begin{IEEEproof}
    The proof of the direct parts appears in App. \ref{appendix:theorem_mmib_mismatched_decoder_rate_direct_proof} and the proof of  ensemble tightness appears in App. \ref{appendix:theorem_mmib_mismatched_decoder_rate_converse_proof}.
    \end{IEEEproof}
}
{
\begin{IEEEproof}
    The proof of the direct parts appears in \cite[App. B]{Dikshtein2023} and the proof of  ensemble tightness appears in \cite[App. C]{Dikshtein2023}.
\end{IEEEproof}
}

The resulting capacity expression of Theorem \ref{theorem:mmib_mismatched_decoder_rate} is proved for the discrete case and is not easily extended to continuous channels. Therefore, in what follows, we provide an achievable random-coding rate based on the \acrfull{gmi} coding scheme \cite{kaplan1993information}. As we shall see, this rate can be extended to continuous channels. Note that the main difference between the LM rate from \Cref{theorem:mmib_mismatched_decoder_rate} and the \acrshort{gmi} rate provided in the following theorem is the relaxation of the marginals equality constraint, i.e., $\pmfQ{\rv{X}} = \pmf{\rv{X}} $.
\begin{theorem} \label{theorem:mmib_mismatched_decoder_gmi_rate}
    The random coding capacity of the \acrshort{dmibc} with mismatched decoder is lower bounded as 
    \begin{equation}
        \begin{aligned}
            &C_V(B) \geq
            && \max_{\pmf{\rv{X}}, \pmf{\rv{Z}|\rv{Y}}}
            && I_{\mathsf{GMI}}(\pmf{\rv{X} \rv{Z}} )  \\
            &&& \text{subject to}
            && I(\rv{Y};\rv{Z}) \leq B,
        \end{aligned}
    \end{equation}
    where
    \begin{equation} \label{eq:mmibdec_Igmi_optimization_problem}
        \begin{aligned}
            &I_{\mathsf{GMI}}(\pmf{\rv{X} \rv{Z} } ) = 
            &&\min_{\pmfT{\rv{X} \rv{Z}}}
            && \kld{ \pmfT{\rv{X}\rv{Z}}}{\pmf{\rv{X}} \times \pmf{\rv{Z}}} \\
            &&& \text{s.t.} 
            && \sum_{x \in \set{X}} \pmfT{\rv{X} \rv{Z}} (x,z) = \pmf{\rv{Z}}(z) \\
            &&&&& \Exp[\pmfT{\rv{X}\rv{Z}}]{V(\rv{Z}|\rv{X})} \geq \Exp[\pmf{\rv{X}\rv{Z}}]{V(\rv{Z}|\rv{X})}.
        \end{aligned}
    \end{equation}
    In addition, the inner minimization problem has the following dual form,
    \begin{equation} \label{eq:mmdec_gmi_dual}
    I_{\mathsf{GMI}}(\pmf{\rv{X} \rv{Z}} ) = \max_{\lambda \geq 0} \mkern-10mu \sum_{(x,z) \in \set{X}\times \set{Z}} \mkern-20mu \pmf{\rv{X}\rv{Z}}(x,z) \log \frac{V(z|x)^{\lambda}}{\sum_{x' } \pmf{\rv{X}}(x') V(z|x')^{\lambda}}.
\end{equation}
\end{theorem}
\ifthenelse{\boolean{full}}
{
\begin{IEEEproof}
    The proof appears in App. \ref{appendix:theorem_mmib_mismatched_decoder_gmi_rate_direct_proof}.
\end{IEEEproof}
}
{
\begin{IEEEproof}
    The proof appears in \cite[App. D]{Dikshtein2023}.
\end{IEEEproof}
}

\subsection{A Computationally Efficient Algorithm}
\input{./sections/mmdec/mmdec_algorithm}

\subsection{Example: A Quaternary Channel}
\input{./sections/mmdec/mmdec_quaternary_channel}

%% file: figures/mmib/block_diagram.tex
\def\dMEnc{1}
\def\dEncCh{2.5}
\def\dChRelay{2}
\def\dRelayDec{3}
\def\dDecHatM{1.25}
\begin{tikzpicture}[every text node part/.style={align=center},thick,scale=0.8, every node/.style={scale=0.8}]
    \coordinate (M) at (0,0);
    \node[draw=black,  minimum height = 1cm] (enc) at (\dMEnc,0) {Channel\\ Encoder}; 
    \node[draw=black,  minimum height = 1cm] (ch) at (\dMEnc+\dEncCh,0) { $\pmf{\rv{Y}| \rv{X}} $};
    \node[draw=black,  minimum height = 1cm] (relay) at (\dMEnc+\dEncCh + \dChRelay,0) {Relay \\ Encoder};

    \node[draw=black,  minimum height = 1cm] (dec) at (\dMEnc+\dEncCh + \dChRelay+\dRelayDec,0) {Mismatched \\Decoder};

    \coordinate (hatM) at (\dMEnc+\dEncCh + \dChRelay+\dRelayDec+\dDecHatM,0);

    \draw[->] (M) node[left] {$\rv{M}$} -- (enc);
    \draw[->] (enc) --  node[above] {$\rv{X}^n(\rv{M})$} (ch);
    \draw[->] (ch) --  node[above] {$\rv{Y}^n$} (relay);
    \draw[->] (relay) -- node[above] {$\rv{Z}^n(\rv{W})$} (dec);
    \draw[->] (dec) -- (hatM) node[right]{$\hat{\rv{M}}$};
\end{tikzpicture}

%% file: sections/mmdec/mmdec_algorithm.tex
In this section, we propose an efficient algorithm to compute $C_V(B)$ by solving  the optimization problem of \Cref{theorem:mmib_mismatched_decoder_rate}.
To this end, it should be noted that  the inner minimization problem of computing $J_V(\rv{X};\rv{Z})$ in  \eqref{eq:mmibdec_Jxz_optimization_problem}  is a convex optimization problem in $\pmfQ{\rv{Z}|\rv{X}}$ since  the mutual information is a convex function of the channel and the constraints on $\pmfQ{\rv{Z}|\rv{X}}$ are linear. Therefore, $J_V(\rv{X};\rv{Z})$ can be efficiently computed using standard convex optimization solvers. 

By contrast, the outer \textit{maximization} problem in \eqref{eq:mmibdec_Jxz_outer_optimization_problem} over $\pmf{\rv{Z}|\rv{Y}}$ is not concave, and solving it requires global optimization methods, e.g., a grid search. Specifically, we propose to initially compute this maximum over a coarse grid of the probability simplex. Then, we perform a refined search of the maximum in a finer grid, only at a local neighborhood of the solution of the coarse maximization. Repeating this refinement procedure in an iterative manner then leads to an improved solution at each step, and a stopping criterion may be a negligible increase in the achievable rate in the last iteration. We nonetheless emphasize that any choice of $\pmf{\rv{Z}|\rv{Y}}$ leads to an achievable lower bound on the rate. Thus, the crucial optimization step for the validity of the solution is the convex minimization step in \eqref{eq:mmibdec_Jxz_optimization_problem}.
In addition, one may also optimize the input distribution $\pmf{\rv{X}}$, again, using search methods. In many practical applications, however, the input distribution is arbitrarily chosen, e.g., as a uniform distribution, which is typically justified by the symmetry of the problem. 


The proposed algorithm involves an alternating maximization step, proposed in \cite{Tishby1999} for the original IB problem, in order to find the optimal test-channel in the case that the decoder is matched. As well known, this algorithm is based on the Blahut-Arimoto \cite{Blahut1972,Arimoto1972}  algorithm and is termed here \acrfull{ibam}. A formal description can be found, e.g., in \cite[Thm. 5]{Tishby1999}. Our main algorithm is detailed in \Cref{alg:mmib}.


\SetKwComment{Comment}{/* }{ */}

\begin{algorithm}[hbt!]
	\DontPrintSemicolon
	\caption{\acrshort{mmib} algorithm }\label{alg:mmib}
	\KwIn{$\pmf{\rv{X}\rv{Y}}, V(z|x), B, \textsf{RES} $}
    $\pmf{\rv{Z}|\rv{Y}}^{opt} = \text{IBAM}( \pmf{\rv{X} \rv{Y}}, C, args)$ \;
	$ \set{P}_{\set{Z}|\set{Y}}  = \{ \pmf{\rv{Z}|\rv{Y}} \colon I(\rv{Y};\rv{Z}) = B \} $\;
    $ \set{P}_{\set{Z}|\set{Y}}^{\text{coarse}} = \set{P}_{\set{Z}|\set{Y}} \cap \textsf{RES} \cdot \field{Z}^{|\set{Z}|} $ \;

    \For{$\pmf{\rv{Z}|\rv{Y}} \in \set{P}_{\set{Z}|\set{Y}}^{\text{coarse}}$}
    {
    Compute $D$ according to \eqref{eq:mmibdec_lm_rate_contraintD} with $\pmf{} = \pmf{\rv{Z}|\rv{Y}} \cdot \pmf{\rv{X}\rv{Y}}$ \;
    $J_V(\rv{X};\rv{Z}) = \min_Q I(\pmf{\rv{X}},  \pmfQ{\rv{Z}|\rv{X}} ) $ \;
    subject to: \;
    $\Exp[\pmf{Q}]{\log V(\rv{Z}|\rv{X})} \geq -D $
    }
    Find $\pmf{\rv{Z}|\rv{Y}}^* \in \set{P}_{\set{Z}|\set{Y}}^{\text{coarse}} $ that maximizes $J_V(\rv{X};\rv{Z})$. \;
	\While{$ \Delta R \geq \epsilon $}{
        $\text{RES} = \text{RES} * \text{FINE}$\;
        $ \set{P}_{\set{Z}|\set{Y}}^{\text{fine}} = \set{P}_{\set{Z}|\set{Y}} \cap \textsf{FINE} \cdot \field{Z}^{|\set{Z}|} \cap \set{B}(\pmf{\rv{Z}|\rv{Y}}^*, 1/ \text{RES}) $ \;
		\For{$\pmf{\rv{Z}|\rv{Y}} \in \set{P}_{\set{Z}|\set{Y}}^{\text{fine}}$}
        {
        Compute $D$ according to \eqref{eq:mmibdec_lm_rate_contraintD} with $\pmf{} = \pmf{\rv{Z}|\rv{Y}} \cdot \pmf{\rv{X}\rv{Y}}$ \;
    $J_V(\rv{X};\rv{Z}) = \min_Q I(\pmf{\rv{X}},  \pmfQ{\rv{Z}|\rv{X}} ) $ \;
    subject to: \;
    $\Exp[\pmf{Q}]{\log V(\rv{Z}|\rv{X})} \geq -D $
        }
        Find $\pmf{\rv{Z}|\rv{Y}}^* \in \set{P}_{\set{Z}|\set{Y}}^{\text{fine}} $ that maximizes $J_V(\rv{X};\rv{Z})$. \;
	}
	\KwOut{$ \pmf{\rv{Z}|\rv{Y}}^*  $, $R$}
\end{algorithm}



%% file: sections/mmdec/mmdec_quaternary_channel.tex
In this section, we demonstrate the result of \Cref{theorem:mmib_mismatched_decoder_rate} and the operation of \Cref{alg:mmib} in a simple setting. Concretely, suppose that the channel from $\rv{X}$ to $\rv{Y}$ is defined by the following conditional \acrshort{pmf},
\begin{equation}
    \pmf{\rv{Y}|\rv{X}}(y|x) = \Prob{\rv{Y}=y|\rv{X}=x} 
    =
    \begin{cases}
        1-\epsilon, & y=x \\
        \frac{\epsilon}{2}, & y=x\pm 1\\
        0, & \text{otherwise},
    \end{cases}
\end{equation}
where all addition and subtraction operations are computed modulo $4$.  The metric $V$ is mismatched, and specifically, is matched to a different channel. This different channel has the same  correct symbol transition probability of $1-\epsilon$, yet the error can go to the other three alternative symbols with probability $\frac{\epsilon}{3}$, i.e.,
\begin{equation}
    \pmfQ{\rv{Y}|\rv{X}}(y|x) = \Prob{\rv{Y}=y|\rv{X}=x} 
    =
    \begin{cases}
        1-\epsilon, & y=x \\
        \frac{\epsilon}{3},  & y \neq x.
    \end{cases}
\end{equation}

\begin{figure}[t]
    \centering
    \input{./figures/mmdec/simulations/mmibdec_quatrenary}
    \caption{Mismatched performance of the Quaternary example.}
    \label{fig:mmdec_quatrenary_simulation}
\end{figure}
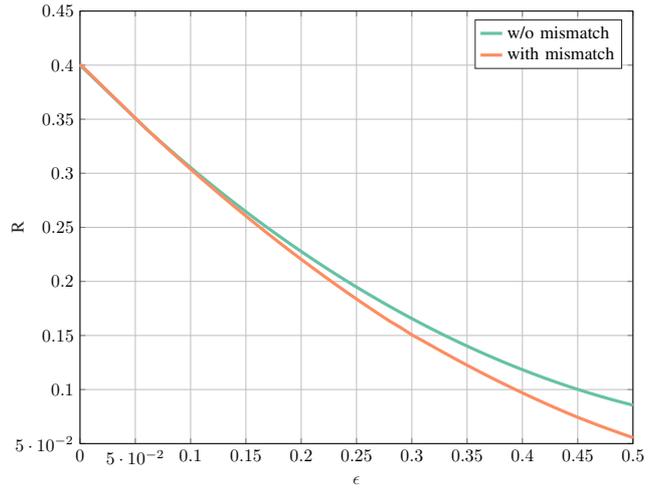

Due to symmetry, we assume that the optimal $\pmf{\rv{X}}$ is uniform and $\pmf{\rv{Z}|\rv{Y}}$ is symmetric. In such case, $\pmf{\rv{Z}|\rv{Y}} (z|y)$ is a modulo-additive channel.
We compare the rates obtained by a matched vs. a mismatched decoder in \Cref{fig:mmdec_quatrenary_simulation}, as a function of $\epsilon$. As might be expected, the difference between the rates vanishes as $\epsilon\to 0$ since the mismatch between the channels becomes milder as  $\epsilon\to 0$.

%% file: figures/mmdec/simulations/mmibdec_quatrenary.tex
%
%
\definecolor{mycolor1}{rgb}{0.40000,0.76078,0.64706}%
\definecolor{mycolor2}{rgb}{0.98824,0.55294,0.38431}%
\begin{tikzpicture}[thick,scale=0.475, every node/.style={scale=1.4}]

\begin{axis}[%
width=6.028in,
height=4.754in,
at={(1.011in,0.642in)},
scale only axis,
xmin=0,
xmax=0.5,
xlabel style={font=\color{white!15!black}},
xlabel={$\epsilon$},
ymin=0.05,
ymax=0.45,
ylabel style={font=\color{white!15!black}},
ylabel={R},
axis background/.style={fill=white},
xmajorgrids,
ymajorgrids,
legend style={legend cell align=left, align=left, draw=white!15!black}
]
\addplot [color=mycolor1, line width=2.5pt]
  table[row sep=crcr]{%
0	0.400011112397723\\
0.06	0.341011598170554\\
0.07	0.331820612555064\\
0.08	0.322806213972453\\
0.09	0.313966125750779\\
0.1	0.305298195629291\\
0.11	0.296800388739585\\
0.12	0.288470781186017\\
0.13	0.280307554166761\\
0.14	0.272308988583851\\
0.15	0.264473460096579\\
0.16	0.256799434577858\\
0.17	0.2492854639377\\
0.18	0.241930182281986\\
0.19	0.234732302378158\\
0.2	0.227690612402565\\
0.21	0.220803972946889\\
0.22	0.214071314263479\\
0.23	0.207491633731532\\
0.24	0.20106399352797\\
0.25	0.194787518488503\\
0.26	0.188661394145923\\
0.27	0.182684864933983\\
0.28	0.176857232546444\\
0.29	0.171177854441968\\
0.3	0.165646142486538\\
0.31	0.160261561725956\\
0.32	0.155023629281848\\
0.33	0.149931913365308\\
0.34	0.144986032403034\\
0.35	0.140185654271484\\
0.36	0.135530495635111\\
0.37	0.131020321385378\\
0.38	0.126654944177738\\
0.39	0.122434224064283\\
0.4	0.118358068220262\\
0.41	0.114426430763122\\
0.42	0.110639312663187\\
0.43	0.106996761745551\\
0.44	0.103498872783176\\
0.45	0.100145787681667\\
0.46	0.0969376957566163\\
0.47	0.0938748341048729\\
0.48	0.0909574880715773\\
0.49	0.0881859918152644\\
0.5	0.0855607289738617\\
};
\addlegendentry{w/o mismatch}

\addplot [color=mycolor2, line width=2.5pt]
  table[row sep=crcr]{%
0	0.400979081612065\\
0.06	0.340997446773845\\
0.07	0.331498993884016\\
0.08	0.322140191490632\\
0.09	0.312920128693178\\
0.1	0.303837944180931\\
0.11	0.294892824905696\\
0.12	0.286084004502954\\
0.13	0.277410761885529\\
0.14	0.268872419948013\\
0.15	0.26046834439456\\
0.16	0.252197942483726\\
0.17	0.244060662258185\\
0.18	0.236055991412355\\
0.19	0.228183456472487\\
0.2	0.220442622026344\\
0.21	0.212833090004138\\
0.22	0.205354499007712\\
0.23	0.198006523732346\\
0.24	0.19078887446284\\
0.25	0.183701296612055\\
0.26	0.176743570230217\\
0.27	0.169915509872727\\
0.28	0.163216964040565\\
0.29	0.157224434032844\\
0.3	0.150578985047797\\
0.31	0.145059364378649\\
0.32	0.139255175477558\\
0.33	0.133563916339868\\
0.34	0.127960902643263\\
0.35	0.122495598609603\\
0.36	0.117143422805635\\
0.37	0.111910006358598\\
0.38	0.106794946857787\\
0.39	0.101797345377633\\
0.4	0.0969191365708269\\
0.41	0.0921544592004682\\
0.42	0.0875029538693448\\
0.43	0.0829648856856541\\
0.44	0.0785734515189523\\
0.45	0.0743173619178513\\
0.46	0.0703559049399498\\
0.47	0.0664205182435852\\
0.48	0.0627273731396148\\
0.49	0.0590489265557268\\
0.5	0.0555064262726841\\
};
\addlegendentry{with mismatch}

\end{axis}
\end{tikzpicture}%

%% file: sections/mmdec_gen.tex

In this section, we modify the bound of the previous section to continuous alphabet Gaussian channels. As mentioned, the discrete alphabet assumption is crucial to the derivation of the LM rate in \Cref{theorem:mmib_mismatched_decoder_rate}. Indeed, in the standard point-to-point communication setting, without mismatch, coding theorems for continuous alphabets are obtained by taking the limit of fine quantization of the continuous inputs and outputs of the channel. Unfortunately, this technique is not applicable in the mismatched case, since quantization of the output changes the decoder which should be fixed by assumption, and it becomes very challenging to track and evaluate the impact on the final result. For this purpose, we have also derived the corresponding  GMI rate in \Cref{theorem:mmib_mismatched_decoder_gmi_rate}, as a lower bound, which, as we shall see, is amenable to modification from discrete alphabet channels to continuous alphabet channels.

From a technical notation perspective, we first replace all \acrshort{pmf}s to the form of probability densities. Furthermore, the decoding metric $V(z|x)$ will also be defined on $\field{R} \times \field{R}$. In addition, we add a constraint on the input of the channel, otherwise, as happens in most continuous channels, the capacity may be unbounded. Thus, every transmitted sequence $x^n$ must satisfy
   $ \frac{1}{n} \sum_{i=1}^n c(x_i) \leq \Gamma$,
for some cost function $c(x)$ and threshold $\Gamma$. Usually, we take $c(x) = x^2 $ and $\Gamma$, both represent a power constraint and the maximum permitted power per-symbol.

\begin{theorem} \label{theorem:genmmib_mismatched_decoder_rate}
    For any continuous oblivious relay channel with mismatched decoding metric $V$, input cost function $c(\cdot)$ and input cost threshold $\Gamma$, the random coding error probability vanishes for rate $R $ that satisfies:
    \begin{equation}
        \begin{aligned}
            &R \leq 
            && \max_{f_{\rv{X} \rv{Z}}, f_{\rv{Z}|\rv{Y}}}
            && I_{\text{GMI}}(f_{\rv{X} \rv{Z}}) \\
            &&& \text{s.t.}
            && I(\rv{Y};\rv{Z}) \leq B,
            & \Exp{c(\rv{X})} \leq \Gamma.
        \end{aligned}
    \end{equation}
    where
    \begin{equation} \label{eq:mmibdec_Igmi_optimization_problem}
        I_{\mathsf{GMI}}(\pmf{\rv{X} \rv{Z}} ) = \max_{\lambda \geq 0} \mkern-5mu \int \mkern-10mu \mathrm{d}x \mathrm{d}z f_{\rv{X}\rv{Z}}(x,z) \log \frac{V(z|x)^{\lambda}}{\int  \pmf{\rv{X}}(x') V(z|x')^{\lambda} \mathrm{d}x'}.
    \end{equation}
\end{theorem}

\begin{IEEEproof}
    The dual expression from \eqref{eq:mmdec_gmi_dual} can also be derived directly (rather than deriving the dual optimization problem as shown in 
    \ifthenelse{\boolean{full}}{
    \Cref{appendix:theorem_mmib_mismatched_decoder_gmi_rate_direct_proof})
    }
    {\cite[App. ?]{Dikshtein2023}})
     using a similar analysis to that of Gallager \cite{Gallager1968} for maximum-likelihood decoding. The former involves replacing summations with integrals and using a standard expurgation argument to construct a sub-codebook with feasible codewords from the randomly generated codebook.
\end{IEEEproof}



\subsection{Example: Fading Channel}
\input{./sections/mmdec/mmdec_fading_channel}

%% file: sections/mmdec/mmdec_fading_channel.tex
We next exemplify our result on a fading channel, a fundamental  wireless communication channel model. As common, we assume that the channel is complex-valued, and the additive noise is  \acrfull{cscg}. Specifically, we consider a memoryless time-varying fast-fading model of the form
    $\rv{Y}_i = \rv{H}_i \rv{X}_i + \rv{N}_i$,
where $\rv{X}_i \in \field{C}$ is the input, $\rv{N}_i \in \field{C}$ is additive noise, and $\rv{H}_i \in \field{C}$ is a fading coefficient. We assume that $\{\rv{N}_i \}_{i=1}^n$ are \acrshort{iid} distributed according to  $\mathcal{CN} (0,\sigma^2)$, and that $\rv{H}_i$ is an \acrshort{iid} sequence with density function $S_{\rv{H}}$.

\subsubsection{Perfect Channel Knowledge}
If each random realization $\rv{H}_i = h_i$ is  perfectly known at the decoder, then due to Gaussianity of the noise and the optimality of Gaussian compression assuming Gaussian input distribution, the optimal decoding rule would be the following weighted version of the nearest-neighbor rule:
\begin{equation}
    \hat{m} = \argmin_{ j =1,\dots, M} \sum_{i=1}^n |z_i - h_i x_i^{(j)}|^2.
\end{equation}
Similarly, under a power constraint $\Exp{|\rv{X}|^2} \leq \Gamma$ (i.e. $c(x) = |x|^2$) and  a Gaussian input distribution assumption, the optimal rate is achieved using a Gaussian test channel from $\rv{Y}$ to $\rv{Z}$, and is given by
\begin{equation}
    R_{\text{IB}}^{CG}(\Gamma, \sigma^2, S_{\rv{H}}) =  \Exp { \log \frac{|\rv{H}|^2 \Gamma + \sigma^2 + q} {\sigma^2 + q} }.
\end{equation}
Evidently, in a fast-fading channel, it is unrealistic to assume that the decoding is matched, since this requires perfect knowledge of $\rv{H}_i$ at any time point. 

\subsubsection{Imperfect Channel Knowledge}
As said, assuming Gaussian signaling, the pair $(\rv{X},\rv{Y})$ is jointly Gaussian. In the standard \acrshort{ib} setting without mismatch, the optimal test-channel from $\rv{Y}$ to $\rv{Z}$ is also Gaussian in such case. Therefore, we also adopt this test channel for the mismatched setting, and assume that the channel from $\rv{Y}$ to $\rv{Z}$ is Gaussian, i.e., there exists a $\rv{W} \sim \mathcal{CN}(0,q)$ such that
    $\rv{Z} = \rv{Y} + \rv{W} = \rv{H} \rv{X} + \rv{N} + \rv{W}$. 
The value of $q$ is determined as the solution of the mutual information constraint equation, i.e.,
\begin{equation}
    B \geq  I(\rv{Y};\rv{Z}) = \Exp[\rv{H}]{ \log \frac{|\rv{H}|^2 \Gamma + \sigma^2 + q }{q}}.
\end{equation}
Let $q^*$ be the solution to the above equation. We use it to find the \acrshort{gmi} rate of \Cref{theorem:mmib_mismatched_decoder_gmi_rate} using the dual form. 

We adopt a simple uncertainty model in which
\begin{equation}
    \rv{H}_i = \hat{\rv{H}}_i + \Delta_i, \quad \Exp{\Delta| \hat{\rv{H}}_i} = 0,
\end{equation}
where $\hat{\rv{H}}_i$ is a possibly-random estimate of $\rv{H}$ known at the decoder, and $\Delta$ represents an unknown conditionally zero-mean error term. We make the simplifying assumption that the pairs $\{(\hat{\rv{H}}_i, \Delta_i) \}_{i=1}^n$ are \acrshort{iid} with respect to $i=1,\dots,n$, and independent of the channel input and noise.

In the case that the joint density function of $(\hat{\rv{H}}_i, \Delta_i)$ is unknown (or even when it is known but difficult to design a corresponding optimal coding scheme), it is natural to apply weighted nearest-neighbor coding
\begin{equation}
    \hat{m} = \argmin_{j = 1,\dots, M} \sum_{i=1}^n |z_i - \hat{h}_i x_i^{(j)} |^2.
\end{equation}
This is a mismatched decoding rule, in the sense that is would be optimal under a model of the form $\rv{Y} = \hat{\rv{H}} \rv{X} + \rv{N}$. The corresponding decoding metric is given by
    $V(x,z) = e^{-|z-\hat{h}x|^2}$.

\begin{theorem} \label{theorem:mmibdec_fading}
    Consider the complex-valued channel fading setup with a known estimate $|\hat{\rv{H}}|$ at the output and a conditionally zero-mean error term $\Delta$. Under \acrshort{iid} random coding with $\rv{X} \sim \mathcal{CN} (0,\Gamma)$, along with weighted nearest-neighbor decoding, the \acrshort{gmi} rate is given by
    \begin{equation}
    \begin{aligned}
            &C(B) =
            && \max_{q}
            && \Exp{ \log \left( 1+ \frac{ |\hat{\rv{H}}|^2 \Gamma}{\Exp{|\Delta|^2 \big|\hat{\rv{H}} } \Gamma + \sigma^2 + q } \right) } \\
            &&& \text{s.t.}
            && \Exp[\rv{H}]{ \log \frac{|\rv{H}|^2  \Gamma + \sigma^2 + q }{q}} \leq B.
        \end{aligned}
    \end{equation}
\end{theorem}

\begin{IEEEproof}
    The proof is omitted here due to lack of space.
\end{IEEEproof}

\begin{figure}[t]
    \centering
    \input{./figures/mmdec/simulations/mmibdec_fading}
    \caption{Mismatched performance of the fading example.}
    \label{fig:mmdec_fading_simulation}
\end{figure}
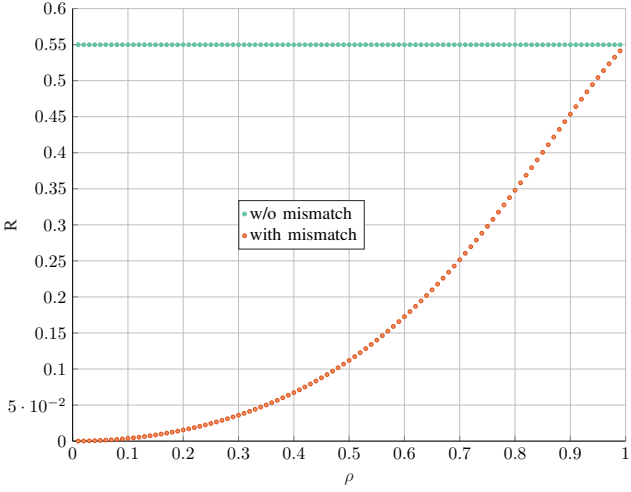

We illustrate \Cref{theorem:mmibdec_fading} using a numerical example. We choose $ \Gamma = \sigma = 1 $ and $\hat{\rv{H}}$ and $\Delta$ to be independent Rayleigh random variables with $\sigma_{\hat{\rv{H}}}^2 = \rho^2$ and $\sigma_{\Delta}^2 = 1-\rho^2$. The results are shown in \Cref{fig:mmdec_fading_simulation}.

%% file: figures/mmdec/simulations/mmibdec_fading.tex
%
%
\definecolor{mycolor1}{rgb}{0.40000,0.76078,0.64706}%
\definecolor{mycolor2}{rgb}{0.98824,0.55294,0.38431}%
\definecolor{mycolor3}{rgb}{0.85000,0.32500,0.09800}%
\begin{tikzpicture}[thick,scale=0.475, every node/.style={scale=1.4}]

\begin{axis}[%
width=6.028in,
height=4.754in,
at={(1.011in,0.642in)},
scale only axis,
xmin=0,
xmax=1,
xlabel style={font=\color{white!15!black}},
xlabel={$\rho$},
ymin=0,
ymax=0.6,
ylabel style={font=\color{white!15!black}},
ylabel={R},
axis background/.style={fill=white},
axis x line*=bottom,
axis y line*=left,
xmajorgrids,
ymajorgrids,
legend style={legend cell align=left, align=left, draw=white!15!black,at={(0.3,0.5)},anchor=west}
]
\addplot[only marks, mark=*, mark options={}, mark size=1.5000pt, draw=mycolor1, fill=mycolor1] table[row sep=crcr]{%
x	y\\
0.01	0.549830075606871\\
0.02	0.549830075606871\\
0.03	0.549830075606871\\
0.04	0.549830075606871\\
0.05	0.549830075606871\\
0.06	0.549830075606871\\
0.07	0.549830075606871\\
0.08	0.549830075606871\\
0.09	0.549830075606871\\
0.1	0.549830075606871\\
0.11	0.549830075606871\\
0.12	0.549830075606871\\
0.13	0.549830075606871\\
0.14	0.549830075606871\\
0.15	0.549830075606871\\
0.16	0.549830075606871\\
0.17	0.549830075606871\\
0.18	0.549830075606871\\
0.19	0.549830075606871\\
0.2	0.549830075606871\\
0.21	0.549830075606871\\
0.22	0.549830075606871\\
0.23	0.549830075606871\\
0.24	0.549830075606871\\
0.25	0.549830075606871\\
0.26	0.549830075606871\\
0.27	0.549830075606871\\
0.28	0.549830075606871\\
0.29	0.549830075606871\\
0.3	0.549830075606871\\
0.31	0.549830075606871\\
0.32	0.549830075606871\\
0.33	0.549830075606871\\
0.34	0.549830075606871\\
0.35	0.549830075606871\\
0.36	0.549830075606871\\
0.37	0.549830075606871\\
0.38	0.549830075606871\\
0.39	0.549830075606871\\
0.4	0.549830075606871\\
0.41	0.549830075606871\\
0.42	0.549830075606871\\
0.43	0.549830075606871\\
0.44	0.549830075606871\\
0.45	0.549830075606871\\
0.46	0.549830075606871\\
0.47	0.549830075606871\\
0.48	0.549830075606871\\
0.49	0.549830075606871\\
0.5	0.549830075606871\\
0.51	0.549830075606871\\
0.52	0.549830075606871\\
0.53	0.549830075606871\\
0.54	0.549830075606871\\
0.55	0.549830075606871\\
0.56	0.549830075606871\\
0.57	0.549830075606871\\
0.58	0.549830075606871\\
0.59	0.549830075606871\\
0.6	0.549830075606871\\
0.61	0.549830075606871\\
0.62	0.549830075606871\\
0.63	0.549830075606871\\
0.64	0.549830075606871\\
0.65	0.549830075606871\\
0.66	0.549830075606871\\
0.67	0.549830075606871\\
0.68	0.549830075606871\\
0.69	0.549830075606871\\
0.7	0.549830075606871\\
0.71	0.549830075606871\\
0.72	0.549830075606871\\
0.73	0.549830075606871\\
0.74	0.549830075606871\\
0.75	0.549830075606871\\
0.76	0.549830075606871\\
0.77	0.549830075606871\\
0.78	0.549830075606871\\
0.79	0.549830075606871\\
0.8	0.549830075606871\\
0.81	0.549830075606871\\
0.82	0.549830075606871\\
0.83	0.549830075606871\\
0.84	0.549830075606871\\
0.85	0.549830075606871\\
0.86	0.549830075606871\\
0.87	0.549830075606871\\
0.88	0.549830075606871\\
0.89	0.549830075606871\\
0.9	0.549830075606871\\
0.91	0.549830075606871\\
0.92	0.549830075606871\\
0.93	0.549830075606871\\
0.94	0.549830075606871\\
0.95	0.549830075606871\\
0.96	0.549830075606871\\
0.97	0.549830075606871\\
0.98	0.549830075606871\\
0.99	0.549830075606871\\
};
\addlegendentry{w/o mismatch}

\addplot[only marks, mark=*, mark options={}, mark size=1.5000pt, draw=mycolor3, fill=mycolor2] table[row sep=crcr]{%
x	y\\
0.01	3.73293858713113e-05\\
0.02	0.000149352313487786\\
0.03	0.000336173113232091\\
0.04	0.000597965736838855\\
0.05	0.000934973858467773\\
0.06	0.00134751101506726\\
0.07	0.00183596078486903\\
0.08	0.00240077700251136\\
0.09	0.00304248400891988\\
0.1	0.00376167693369855\\
0.11	0.00455902200737326\\
0.12	0.00543525690040816\\
0.13	0.00639119108545657\\
0.14	0.00742770621881196\\
0.15	0.00854575653650821\\
0.16	0.00974636925994197\\
0.17	0.0110306450052786\\
0.18	0.0123997581902438\\
0.19	0.0138549574311763\\
0.2	0.0153975659224459\\
0.21	0.0170289817894913\\
0.22	0.0187506784058179\\
0.23	0.0205642046633057\\
0.24	0.0224711851840973\\
0.25	0.0244733204611737\\
0.26	0.0265723869134681\\
0.27	0.0287702368400004\\
0.28	0.0310687982560465\\
0.29	0.0334700745927733\\
0.3	0.035976144240055\\
0.31	0.0385891599103588\\
0.32	0.0413113477996062\\
0.33	0.0441450065188176\\
0.34	0.0470925057680754\\
0.35	0.0501562847219486\\
0.36	0.0533388500929495\\
0.37	0.0566427738368817\\
0.38	0.0600706904610695\\
0.39	0.0636252938934227\\
0.4	0.067309333867126\\
0.41	0.0711256117724143\\
0.42	0.075076975923446\\
0.43	0.0791663161847237\\
0.44	0.0833965578978439\\
0.45	0.0877706550456288\\
0.46	0.0922915825869293\\
0.47	0.0969623278916278\\
0.48	0.101785881201677\\
0.49	0.106765225040433\\
0.5	0.111903322489174\\
0.51	0.117203104246608\\
0.52	0.122667454384477\\
0.53	0.128299194710225\\
0.54	0.134101067646154\\
0.55	0.140075717533855\\
0.56	0.146225670273027\\
0.57	0.15255331120538\\
0.58	0.159060861157389\\
0.59	0.16575035056045\\
0.6	0.172623591573862\\
0.61	0.179682148145261\\
0.62	0.186927303955115\\
0.63	0.194360028206935\\
0.64	0.201980939243464\\
0.65	0.209790265991608\\
0.66	0.21778780726579\\
0.67	0.225972888991092\\
0.68	0.234344319444449\\
0.69	0.24290034265464\\
0.7	0.251638590150273\\
0.71	0.260556031299487\\
0.72	0.269648922546017\\
0.73	0.278912755913415\\
0.74	0.288342207222496\\
0.75	0.297931084546052\\
0.76	0.307672277508744\\
0.77	0.317557708128005\\
0.78	0.327578283982119\\
0.79	0.337723854582771\\
0.8	0.347983171918833\\
0.81	0.358343856223237\\
0.82	0.368792368092169\\
0.83	0.379313988151681\\
0.84	0.389892805516864\\
0.85	0.400511716318309\\
0.86	0.411152433574749\\
0.87	0.421795509664331\\
0.88	0.432420372585079\\
0.89	0.443005377092921\\
0.9	0.453527871659291\\
0.91	0.463964281996652\\
0.92	0.474290211657691\\
0.93	0.484480559922533\\
0.94	0.494509656850227\\
0.95	0.50435141499062\\
0.96	0.51397949683784\\
0.97	0.523367496666856\\
0.98	0.532489134942718\\
0.99	0.541318463043218\\
};
\addlegendentry{with mismatch}

\end{axis}

\begin{axis}[%
width=7.778in,
height=5.833in,
at={(0in,0in)},
scale only axis,
xmin=0,
xmax=1,
ymin=0,
ymax=1,
axis line style={draw=none},
ticks=none,
axis x line*=bottom,
axis y line*=left,
legend style={legend cell align=left, align=left, draw=white!15!black}
]
\end{axis}
\end{tikzpicture}%

%% file: sections/mmrib_main_results.tex
\begin{figure}[b]
    \centering
    \input{./figures/mmrib/block_diagram}
    \caption{Oblivious Communication System with Mismatched Relay}
    \label{fig:mmrib_communication_diagram}
\end{figure}
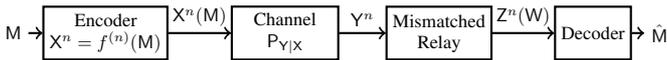

In this section, we consider a different relay model with a mismatch. Specifically, let us consider the 3-node point-to-point communication system with a mismatched relay depicted in \Cref{fig:mmrib_communication_diagram}. In this model, the sender wishes to communicate a message $\rv{M}$ to the receiver with the help of the relay. We consider the \acrfull{dmibc} with mismatched relay $ \langle \rv{X}, \pmf{\rv{Y}|\rv{X}},  \set{Y}, \set{Z}, d_0(y,z) \rangle $ that consists of three finite sets $\set{X}$, $\set{Y}$, $\set{Z}$, a collection of conditional \acrshort{pmf}s $\pmf{\rv{Y}|\rv{X}}$ on $\set{Y}$ (one for each $x$), and an encoding metric $d_0(y,z)$ on $\set{Y} \times \set{Z}$.

A $(2^{nR},2^{nB},n)$ code for the \acrshort{dmibc} with mismatched relay defined in a similar manner to \Cref{section:mmdec} with the following exceptions:
\begin{itemize}
    \item a mismatched relay encoder that assigns an index $\hat{w} \in \set{W}$ to each received sequence $y^n \in \set{Y}^n$ according to $\hat{w} = \argmin_{w \in \set{W}} d_0^n(y^n,z^n(w))$ where
    \begin{equation}
        d_0^n(y^n,z^n) = \frac{1}{n} \sum_{i=1}^n d_0(y_i,z_i),
    \end{equation}
    \item a decoder that assigns an estimate $\hat{m}$ or an error message e to each received representation index $w \in \set{W}$. 
\end{itemize}
We assume that the decoder knows the channel and mismatched relay's codebook.
Furthermore, it is assumed that the message $\rv{M}$ is uniformly distributed over the message set $\set{M}$. 


\begin{remark}
    The mismatched compression problem is somewhat simpler in  case  the mismatch at the relay is the result of a wrong test-channel for compression, that is,  the relay is constrained to joint typicality encoding with $\pmfQ{\rv{Z}|\rv{Y}}$. This is equivalent to a choice of nonoptimal test-channel in the standard \acrshort{ib} problem. In such case, the resulting capacity is given by
    \begin{equation}
        C(B,\pmfQ{\rv{Z}|\rv{Y}}) = 
        \begin{cases}
            \max_{\pmf{\rv{X}}} I(\rv{X};\rv{Z}), & I(\pmf{\rv{Y}}, \pmfQ{\rv{Z}|\rv{Y}}) \leq B \\
            0, &\text{otherwise}
        \end{cases}
    \end{equation}
    where $\pmfQ{\rv{Z}|\rv{X}}(z|x) = \sum_{y \in \set{Y}} \pmfQ{\rv{Z}|\rv{Y}}(z|y) \pmf{\rv{Y}|\rv{X}}(y|x)$.
    Note the  difference between this setting and the mismatched relay compression setting. In the standard \acrshort{ib} problem, a test-channel $\pmf{\rv{Z}|\rv{Y}}$ is optimized, and a random compression codebook is generated according to $\pmf{\rv{Z}}$. Given this codebook, the compressed index is chosen based on joint typicality encoding with the given $\pmf{\rv{Y}\rv{Z}}$. By contrast, in the mismatched relay setting, the compressed codeword is chosen according to the given fixed (mismatched) metric, which is not necessarily matched to $\pmf{\rv{Y}\rv{Z}}$.
\end{remark}

Our main result for this setting is stated in the following theorem.
\begin{theorem} \label{theorem:mmib_mismatched_relay_rate}
    The capacity of the  mismatched relay channel is lower bounded as:
    \begin{equation}
        \begin{aligned}
            &C(B) \geq
            && \max_{\pmf{\rv{X}}, \pmf{\rv{Z}}}
            && \min_{Q_{\rv{Y}\rv{Z}} \in \set{Q}}
            && I(\rv{X};\rv{Z} ) 
        \end{aligned}
    \end{equation}
    where 
        $\set{Q} = \argmin_{\pmfQ{\rv{Y} \rv{Z}} \in \set{D}} \sum_{y \in \set{Y}} \sum_{z \in \set{Z}} \pmfQ{\rv{Y}\rv{Z}} (y,z) d_0(y,z)$
    and
    \begin{equation}
        \set{D} = \left\{
        \substack{\pmfQ{\rv{Y}\rv{Z}} \colon 
            \pmfQ{\rv{Y}} = \pmf{\rv{Y}},
            \pmfQ{\rv{Z}} = \pmf{\rv{Z}}, \\
            \pmfQ{\rv{X} \rv{Z}}(x,z) = \sum_{y \in \set{Y}} \pmf{\rv{X}|\rv{Y}}(x|y) \pmfQ{\rv{Y}\rv{Z}}(y,z) , 
             I(\rv{Y};\rv{Z}) \leq B
             }\right\}.
    \end{equation}
\end{theorem}
\ifthenelse{\boolean{full}}
{
\begin{IEEEproof}
    The proof appears  in App. \ref{appendix:theorem_mmib_mismatched_relay_rate_direct_proof}.
\end{IEEEproof}
}
{
\begin{IEEEproof}
    The proof appears  in \cite[App. E]{Dikshtein2023}.
\end{IEEEproof}
}

%% file: figures/mmrib/block_diagram.tex
\def\dMEnc{1.5}
\def\dEncCh{3.5}
\def\dChRelay{3}
\def\dRelayDec{3}
\def\dDecHatM{1}
\begin{tikzpicture}[every text node part/.style={align=center},thick,scale=0.675, every node/.style={scale=0.7}]
    \coordinate (M) at (0,0);
    \node[draw=black, minimum width=2cm, minimum height = 1cm] (enc) at (\dMEnc,0) {Encoder \\ $\rv{X}^n = f^{(n)}(\rv{M}) $};
    \node[draw=black, minimum width=2cm, minimum height = 1cm] (ch) at (\dMEnc+\dEncCh,0) {Channel\\ $\pmf{\rv{Y}| \rv{X}} $};
    \node[draw=black,  minimum height = 1cm] (relay) at (\dMEnc+\dEncCh + \dChRelay,0) {Mismatched \\ Relay };

    \node[draw=black,  minimum height = 1cm] (dec) at (\dMEnc+\dEncCh + \dChRelay+\dRelayDec,0) {Decoder};

    \coordinate (hatM) at (\dMEnc+\dEncCh + \dChRelay+\dRelayDec+\dDecHatM,0);

    \draw[->] (M) node[left] {$\rv{M}$} -- (enc);
    \draw[->] (enc) --  node[above] {$\rv{X}^n(\rv{M}) $} (ch);
    \draw[->] (ch) --  node[above] {$\rv{Y}^n$} (relay);
    \draw[->] (relay) -- node[above] {$ \rv{Z}^n (\rv{W})$} (dec);
    \draw[->] (dec) -- (hatM) node[right]{$\hat{\rv{M}}$};
\end{tikzpicture}

%% file: sections/compound.tex
\begin{figure}[b]
    \centering
    \input{./figures/dmibc/compound_block_diagram}
    \caption{Compound Oblivious Communication System}
    \label{fig:dmibc_compound_diagram}
\end{figure}

The compound \acrshort{ib} channel is a \acrshort{dmibc} with state, where the channel is selected arbitrary from a set of possible \acrshort{dmc}s and fixed throughout the transmission block as depicted in \autoref{fig:dmibc_compound_diagram}. It models communication in the presence of uncertainty about channel statistics. For clarity of notation, we use the equivalent definition of a compound \acrshort{ib} channel as consisting of a set of \acrshort{dmc}s $\langle \set{X} , \pmf{\rv{Y}_{s}|\rv{X}}, \set{Y} \rangle$ where $y_s \in \set{Y}$ for every state $s$ in the finite set $\set{S}$. The state $s$ remains the same throughout the transmission block, i.e., $\pmf{\rv{Y}^n|\rv{X}^n \rv{S}^n} (y^n|x^s,s^n) = \prod_{i=1}^n \pmf{\rv{Y}_s|\rv{X}} (y_i|x_i) = \prod_{i=1}^n \pmf{\rv{Y}|\rv{X} \rv{S}} (y_i|x_i,s) $.

Consider first the case where the state is not available at either the encoder, the relay, or the decoder. In addition we assume that the relay is oblivious of the transmitter's codebook. The definition of $(2^{nR}, 2^{nB},n)$ code under this assumption is the same as for the \acrshort{dmibc}. The average probability of error, however, is defined as
\begin{equation}
    P_e^{(n)} = \max_{s \in \set{S}} \Prob{\hat{\rv{M}} \neq \rv{M} | s \text{ is the selected channel state}}.
\end{equation}
Achievability and the capacity are also defined as for the \acrshort{dmibc}.

\begin{theorem}\label{theorem:compound_ib_rate}
    The capacity of the compound \acrshort{ib} channel $\langle \set{X} , \{ \pmf{\rv{Y}_{s}|\rv{X}}, s \in \set{S}\}, \set{Y} \rangle$ with no state information available at either the encoder, relay or the decoder is
    \begin{equation}
        \begin{aligned}
            &C_{CIB}(B) = 
            &&\max_{\pmf{\rv{X}}, \pmf{\rv{Z}|\rv{Y}\rv{S}}} 
            &&\min_{s \in \set{S}}  I(\rv{X};\rv{Z}_s) \\
            &&& \text{subject to}
            &&  \max_{s \in \set{S}} I(\rv{Y};\rv{Z}_s) \leq B
        \end{aligned}
    \end{equation}
\end{theorem}

\begin{remark}
    The achievable rate of \autoref{theorem:compound_ib_rate} does not increases when the state $s$ is available at the relay or the decoder.
\end{remark}
\begin{proof}
    The proof is relegated to App. \ref{appendix:theorem_compound_ib_rate_proof}
\end{proof}

\begin{remark}
    Following similar steps as in the proof of \autoref{theorem:compound_ib_rate} one can show that when the state $s$ is available to the encoder, the achievable rate is given by
    \begin{equation}
        \begin{aligned}
            &C_{CIB}(B) = 
            &&\min_{s \in \set{S}} \max_{\pmf{\rv{X}}. \pmf{\rv{Z}|\rv{Y}\rv{S}}} 
            &&I(\rv{X};\rv{Z}_s) \\
            &&& \text{subject to}
            &&  \max_{s \in \set{S}} I(\rv{Y};\rv{Z}_s) \leq B
        \end{aligned}
    \end{equation}
    Furthermore, due to the max-min inequality \cite[Ch. 5.4]{boyd2004} this rate is no lower then when the encoder does not  knows the state $s$, as in \autoref{theorem:compound_ib_rate}.
\end{remark}

%% file: figures/dmibc/compound_block_diagram.tex
\def\dMEnc{1.5}
\def\dEncCh{4}
\def\dEncDec{4}
\def\dChRelay{3}
\def\dRelayDec{3}
\def\dDecHatM{1.5}
\def\dYstate{2}
\begin{tikzpicture}[every text node part/.style={align=center}]
    \coordinate (M) at (0,0);
    \node[draw=black, minimum width=2cm, minimum height = 1cm] (enc) at (\dMEnc,0) {Encoder \\ $\rv{X}^n = f^{(n)}(\rv{M}) $};
    \node[draw=black, minimum width=2cm, minimum height = 1cm] (ch) at (\dMEnc+\dEncCh,0) {$\pmf{\rv{Y}_s| \rv{X}} $};

    \node[draw=black, minimum width=1cm,minimum height = 1cm] (state) at (\dMEnc+\dEncCh,\dYstate) {State\\ $s$};
    \node[draw=black,  minimum height = 1cm] (relay) at (\dMEnc+\dEncCh + \dChRelay,0) {Relay \\ $\rv{W} = g^{(n)}(\rv{Y}^n)   $ };

    \node[draw=black,  minimum height = 1cm] (dec) at (\dMEnc+\dEncCh + \dChRelay+\dRelayDec,0) {Decoder};
    \coordinate (hatM) at (\dMEnc+\dEncCh + \dChRelay + \dRelayDec+\dDecHatM,0);
    
    \draw[->] (M) node[left] {$\rv{M}$} -- (enc);
    \draw[->] (enc) --  node[above] {$\rv{X}^n$} (ch);
    \draw[->] (state) -- (ch);
    \draw[->] (ch) --  node[above] {$\rv{Y}^n$} (relay);
    \draw[->] (relay) -- node[above] {$\rv{W}$} (dec); 
    \draw[->] (dec) -- (hatM) node[right]{$\hat{\rv{M}}$};
\end{tikzpicture}

%% file: sections/ibcsi.tex
\begin{figure}[b]
    \centering
    \input{./figures/dmibc/dmibc_si_block_diagram}
    \caption{Point to Point communication system with a relay and side information.}
    \label{fig:dmibc_si_block_diagram}
\end{figure}

Consider the point-to-point communication system with state depicted in \autoref{fig:dmibc_si_block_diagram}, termed as \textit{the information bottleneck channel with side information}. The sender wishes to communicate a message $m \in \{1:2^{nR} \} = \set{M} $ over a channel with state to the receiver with possible side information about the state sequence $s^n$ available at the encoder and/or the decoder but not at the relay. We consider a discrete memoryless channel with state model $\langle \set{X} \times \set{S}, \pmf{\rv{Y}|\rv{X}\rv{S}}, \pmf{\rv{S}}, \set{Y} \rangle$ that consists of a finite input alphabet $\set{X}$, a finite output alphabet $\set{Y}$, a finite state alphabet $\set{S}$, and a collection of conditional \acrshort{pmf}s $\pmf{\rv{Y}|\rv{X}\rv{S}}$ on $\set{Y}$. The channel is memoryless in the sense that, without feedback, $\pmf{\rv{Y}^n|\rv{X}^n \rv{S}^n} (y^n|x^n,s^n)  =  \prod_{i=1}^n \pmf{\rv{Y}|\rv{X}\rv{S}} (y|x,s)$. The state sequence $(\rv{S}_1,\rv{S}_2,\dots)$ is \acrshort{iid} with $\rv{S}_i \sim \pmf{\rv{S}}$.

\subsection{State Information Available at the Decoder with Oblivious Relay}
\begin{theorem}\label{theoerem:ibcsi_decoder_oblivious_relay_achievable_rate}
    The capacity of the \acrshort{ib} channel with \acrshort{dm} state $ \pmf{\rv{Y}|\rv{X} \rv{S}} \cdot \pmf{\rv{S}}$, and oblivious relay, when the state information is available non-causally only at the decoder is given by
    \begin{equation}
        \begin{aligned}
            &C(B) = 
            &&\max_{\pmf{\rv{X}}, \pmf{\rv{Z}|\rv{Y}}}  
            &&I(\rv{X};\rv{Z}|\rv{S}) \\
            &&& \text{subject to}
            &&  \pmf{\rv{Y}}(y) = \sum_{(x,s)\in \set{X}\times \set{S}} \pmf{\rv{Y}|\rv{X}\rv{S}}(y|x,s) \pmf{\rv{X}}(x) \pmf{\rv{S}}(s), \\
            &&&&& I(\rv{Y};\rv{Z}) \leq B,
        \end{aligned}
    \end{equation}
    where $|\set{Z}| \leq ???$.
\end{theorem}
\begin{proof}
    The proof is relegated to App. \ref{appendix:theorem_ibcsi_decoder_oblivious_relay_rate_proof}.
\end{proof}
\begin{remark}
    This result is in accordance with \cite[Ch. 7.4]{gamal2011}, and it reduces to \cite[Eq. 7.2]{gamal2011} when taking $ B \rightarrow \infty $.
\end{remark}

\subsection{State Information Available at the Decoder with Decoding Relay}
Here we assume that the relay knows the codebook.
\begin{theorem}\label{theoerem:ibcsi_decoder_decoding_relay_achievable_rate}
    The capacity of the \acrshort{ib} channel with \acrshort{dm} state $ \pmf{\rv{Y}|\rv{X} \rv{S}} \cdot \pmf{\rv{S}}$ when the state information is available non-causally only at the decoder and the relay knows the codebook is
    \begin{equation}
        \begin{aligned}
            &C (B) = 
            &&\max_{\pmf{\rv{U}}, \pmf{\rv{X}|\rv{U}}, \pmf{\rv{Z}|\rv{Y}}}  
            && I(\rv{U};\rv{Y})+ I(\rv{X};\rv{Z}|\rv{U},\rv{S}) \\
            &&& \text{subject to}
            &&  \pmf{\rv{Y}}(y) = \sum_{(x,s)\in \set{X}\times \set{S}} \pmf{\rv{Y}|\rv{X}\rv{S}}(y|x,s) \pmf{\rv{X}}(x) \pmf{\rv{S}}(s), \\
            &&&&& I(\rv{U};\rv{Y})+I(\rv{Y};\rv{Z}|\rv{U}) \leq B,
        \end{aligned}
    \end{equation}
    where $|\set{U}| \leq ???$, and $|\set{Z}| \leq ???$.
\end{theorem}
\begin{proof}
    The proof is relegated to App. \ref{appendix:theorem_ibcsi_decoder_decoding_relay_rate_proof}.
\end{proof}

\subsection{Example: QAM}
\input{./sections/qam}

%% file: figures/dmibc/dmibc_si_block_diagram.tex
\def\dMEnc{1.5}
\def\dEncCh{4}
\def\dEncDec{4}
\def\dChRelay{3}
\def\dRelayDec{3}
\def\dDecHatM{1.5}
\def\dYstate{2}
\begin{tikzpicture}[every text node part/.style={align=center}]
    \coordinate (M) at (0,0);
    \node[draw=black, minimum width=2cm, minimum height = 1cm] (enc) at (\dMEnc,0) {Encoder \\ $\rv{X}^n = f^{(n)}(\rv{M}) $};
    \node[draw=black, minimum width=2cm, minimum height = 1cm] (ch) at (\dMEnc+\dEncCh,0) {$\pmf{\rv{Y}| \rv{X} \rv{S}} $};

    \node[draw=black, minimum width=1cm,minimum height = 1cm] (state) at (\dMEnc+\dEncCh,\dYstate) {State\\ $\pmf{\rv{S}}$};
    \node[draw=black,  minimum height = 1cm] (relay) at (\dMEnc+\dEncCh + \dChRelay,0) {Relay \\ $\rv{W} = g^{(n)}(\rv{Y}^n)   $ };

    \node[draw=black,  minimum height = 1cm] (dec) at (\dMEnc+\dEncCh + \dChRelay+\dRelayDec,0) {Decoder};
    \coordinate (hatM) at (\dMEnc+\dEncCh + \dChRelay + \dRelayDec+\dDecHatM,0);
    
    \draw[->] (M) node[left] {$\rv{M}$} -- (enc);
    \draw[->] (enc) --  node[above] {$\rv{X}_i$} (ch);
    \draw[->] (state) -- node [right] {$\rv{S}_i$} (ch);
    \draw[->] (ch) --  node[above] {$\rv{Y}_i$} (relay);
    \draw[->] (relay) -- node[above] {$\rv{W}$} (dec); 
    \draw[->] (state) -- node [above] {$\rv{S}^n$} (state-|dec) -- (dec);
    \draw[->] (dec) -- (hatM) node[right]{$\hat{\rv{M}}$};
\end{tikzpicture}

%% file: sections/qam.tex
Consider the following non-oblivious example. The transmitter transmits a $1024$-QAM constellation. The relay is restricted to decode only the $4$-QAM messages. The encoding and decoding regions are illustrated in \autoref{fig:ibcsi_decoder_decoding_relay_1024qam_examle}.

The relay sends the respective quadrant of the received message (brown dot in the example) and compresses the error vector ($\Vec{c}$ in the example) that corresponds to the quadrant central point.

\begin{figure}[t]
    \centering
    \input{figures/nonoblivious/constellation}
    \caption{1024 QAM constellation with relay decoding regions in red, green, blue and yellow, corresponding to a $2$ bits decoding.}
    \label{fig:ibcsi_decoder_decoding_relay_1024qam_examle}
\end{figure}

%% file: figures/nonoblivious/constellation.tex
\tikzset{
    cpoint/.style = {circle,fill, inner sep=2pt, outer sep = 0pt, node contents ={}}
}
\begin{tikzpicture}[scale=0.5]
\draw[->]  (-16.8,0) -- (16.8,0) node[right] {$\Re$};
\draw[->]  (0,-16.4) -- (0,16.4) node[above] {$\Im$};
\foreach \i in {-16,..., 15}
    \foreach \j in {-16,..., 15}
    {
    \node at (\i+0.5,\j + 0.5) [cpoint];
    }

\draw[fill=red,opacity=0.2] (0,0)--(0,16)--(16,16)-- (16,0) -- cycle;

\draw[fill=green,opacity=0.2] (0,0)--(0,-16)--(16,-16)-- (16,0) -- cycle;

\draw[fill=blue,opacity=0.2] (0,0)--(0,-16)--(-16,-16)-- (-16,0) -- cycle;

\draw[fill=yellow,opacity=0.2] (0,0)--(0,16)--(-16,16)-- (-16,0) -- cycle;

\node[circle,fill=red, inner sep=2pt, outer sep=0pt] at (8,8) {};

\node[circle,fill=green, inner sep=2pt, outer sep=0pt] at (8,-8) {};

\node[circle,fill=blue, inner sep=2pt, outer sep=0pt] at (-8,-8) {};

\node[circle,fill=yellow, inner sep=2pt, outer sep=0pt] at (-8,8) {};

\node[circle,fill=brown, inner sep=2pt, outer sep=0pt] at (13,3) {};

\draw[very thick,->,green] (8,8) -- node[above,sloped]{$\Vec{c}$} (13,3);

\end{tikzpicture}

%% file: sections/conclusions.tex
We considered the problem of reliable communication in a point-to-point oblivious-relay communication system with a mismatch. In particular, we considered mismatch at the relay or at the decoder. We have established ensemble tight achievable rates and their dual representations. We further specialized those results to particular instances: the quaternary channel and the fading channel. We proposed an alternating algorithm to find those rates.

For future work, it would be interesting to consider converse bounds to this problem, e.g., using the methods described in \cite{Scarlett2020} and \cite{Somekh2022}. Alternatively, it would be interesting to find relations between the mismatch capacity of the channel 
$\pmf{\rv{Y}|\rv{X}}$ to that of the entire channel $\pmf{\rv{Z}|\rv{X}}$. Another possibility is to generalize the results to a  state-dependent channel 
\cite{Xu2021b, Xu2021}, where the relay knows the state sequence (which may also be assumed to be i.i.d.), and add a description of the state as part of its message, in an effort to  provide the receiver with channel state information, thus aiding its decoding performance.



%% file: sections/notations.tex
Throughout the paper, random variables are denoted using a sans-serif font, e.g., $ \rv{X} $, their realizations are denoted by the respective lower-case letters, e.g., $ x $, and their alphabets are denoted by the respective calligraphic letters, e.g., $ \mathcal{X} $. Let $ \mathcal{X}^n $ stand for the set of all $ n $-tuples of elements from $ \mathcal{X} $. An element from $ \mathcal{X}^n $ is denoted by $ x^n = (x_1,x_2,\dots , x_n) $.
The vector notation $\bv{x} = x^n$ will be frequently used for simplicity of presentation. 
The cardinality of a finite set, say $ \mathcal{X}$, is denoted by $ |\mathcal{X}|$. The \acrfull{pmf} of $ \rv{X} $, the joint \acrshort{pmf} of $ \rv{X} $ and $ \rv{Y} $, and the conditional \acrshort{pmf} of $ \rv{X} $ given $ \rv{Y} $ are denoted by $ \pmf{\rv{X}} $, $ \pmf{\rv{X} \rv{Y}} $, and $ \pmf{\rv{X}|\rv{Y}} $, respectively. The expectation of $ \rv{X} $ is denoted by $ \Exp{\rv{X}} $. The probability of an event $ \mathcal{E} $ is denoted as $ \Prob{\mathcal{E}} $. Throughout this paper all logarithms are taken to base 2 unless stated otherwise.

The mutual information between $ \rv{X} $ and $ \rv{Y} $ is defined as
\begin{equation}
	I(\rv{X};\rv{Y}) 
 = \sum_{x \in \mathcal{X}} \sum_{y \in \mathcal{Y}} \pmf{\rv{X}\rv{Y}}(x,y) \log \frac{\pmf{\rv{X}\rv{Y}}(x,y)}{\pmf{\rv{X}}(x) \pmf{
			\rv{Y}}(y)}.
\end{equation}
The \acrfull{kl} divergence between two probability measures $\pmf{}$ and $\pmfQ{}$ is defined as
\begin{equation}
    \kld{\pmf{}}{\pmfQ{}} \triangleq 
    \Exp[\pmf{}]{\log \frac{\pmf{}}{\pmfQ{}}}.
\end{equation}



The empirical distribution of $x^n$ will be denoted by $\epmf{\bv{x}}(a)$, for every $a \in \set{X}$, namely
\begin{equation}
    \epmf{\bv{x}}(a) \triangleq \frac{1}{n} \sum_{i=1}^n \indicator{x_i = a},
\end{equation}
where $\indicator{\cdot}$ is the event indicator function. 

For a pair of random variables $(\rv{X},\rv{Y})$ defined by the marginal and conditional probability distributions $\pmf{\rv{X}}$ and $\pmf{\rv{Y}\rv{X}}$ we define a marginalization operation $\circ$ as follows
\begin{equation}
    \pmf{\rv{Y}} = \pmf{\rv{Y}|\rv{X}} \circ \pmf{\rv{X}} \rightleftharpoons \pmf{\rv{Y}}(y) = \sum_{x \in \set{X}} \pmf{\rv{Y}|\rv{X}} (y|x) \pmf{\rv{X}}(x).
\end{equation}



%% file: sections/mmib/lower_bound.tex
Before proceeding to describe the coding scheme, consider first the mismatched decoding rule \eqref{eq:mmdibc_decoding_rule}. Note that
\begin{align}
    V^n(\bv{z}|\bv{x})
    &=
    \prod_{i=1}^n V(z_i|x_i) \\
    &= 2^{n \sum_{(a,c) \in \set{X} \times \set{Z}} \epmf{\bv{x}\bv{z}}(a,c) \log V(c|a)}.
\end{align}
Therefore, the mismatched decoding rule in \eqref{eq:mmdibc_decoding_rule} is equivalent to
\begin{equation} \label{eq:mmdibc_decoding_rule2}
    \hat{m}(\bv{z}) = \argmax_{m \in \set{M}} \sum_{(a,c) \in \set{X} \times \set{Z}} \epmf{\bv{x}(m) \bv{z}}(a,c) \log V(c|a).
\end{equation}

Consider now the following threshold decoding rule. Fix some $\epsilon > 0$. The decoder declares that message $\tilde{m}$ was sent if it is the unique message that satisfies 
\begin{equation}
    \left(\bv{x}(\tilde{m}), \bv{z} (w) \right) \in \typset{\epsilon, \theta}{n},
\end{equation}
where
\begin{equation}
    \typset{\epsilon, \theta}{n} \triangleq \left\{ (\bv{x},\bv{z}) \colon \mkern-10mu \sum_{(a,c) \in \set{X} \times \set{Z}} \mkern-10mu \epmf{\bv{x}\bv{z}} (a,c) \log V(c|a) \geq \theta- \epsilon \right\}.
\end{equation}
Note that if there exists such $\tilde{m}$ then it is also the solution to the true decoding rule in \eqref{eq:mmdibc_decoding_rule2}, i.e., $\tilde{m} = \hat{m}$ Therefore, the error probability of the mismatched decoder in \eqref{eq:mmdibc_decoding_rule2} is upper bounded by the error probability of the threshold decoder. In particular we have
\begin{align}
    \pmf{c}
    &= \Prob{\hat{\rv{M}} = \rv{M}} \\
    &\geq \Prob{\hat{\rv{M}} = \rv{M}, (\rv{X}^n(\rv{M}),\rv{Z}^n(W)) \in \typset{\epsilon,\theta}{n}} \\
    &= \Prob{\text{threshold decoder is correct}}.
\end{align}
In the following we set
    $\theta = \sum_{(a,c) \in \set{X} \times \set{Z}} \pmf{\rv{X}\rv{Z}} (a,c) \log V(c|a) $.

We use random coding. Let $0 < \epsilon'' < \epsilon' < \epsilon$. Fix a \acrshort{pmf} $\pmf{\rv{X}}$ and a conditional \acrshort{pmf} $\pmf{\rv{Z}|\rv{Y}}$, which induce the marginal \acrshort{pmf}s   $\pmf{\rv{Y}}(y) = \sum_{x \in \set{X}} \pmf{\rv{Y}|\rv{X}}(y|x)  \pmf{\rv{X}} (x)$ and $\pmf{\rv{Z}}(z) = \sum_{y \in \set{Y}} \pmf{\rv{Z}|\rv{Y}}(z|y) \pmf{\rv{Y}}(y)$.
\paragraph{Transmitter's Constant-Composition Codebook Generation}
Let $\pmfQ{\rv{X}_n}$ denote an arbitrary type having the same support as $\pmf{\rv{X}}$, and satisfying $\lVert \pmf{\rv{X}} - \pmfQ{\rv{X}_n} \rVert \leq \frac{1}{n}$.
The codewords are drawn independently from the distribution
\begin{equation}
    \pmfQ{\rv{X}^n}(x^n) = \frac{1}{|\set{T}^n(\pmfQ{\rv{X}_n}) |} \indicator{x^n \in \set{T}^{(n)} (\pmfQ{\rv{X}_n})}.
\end{equation}
That is, each codeword is equiprobable on the set of sequences with empirical distribution $\pmfQ{\rv{X}_n}$.
The codebook is revealed to the encoder and the decoder but not the relay.

\paragraph{Relay's Codebook Generation and Compression}
Generate a random codebook $ \set{C}_R = \{ z^n(w) \} $, $w \in [1:2^{nB}]$ according to 
$\pmf{\rv{Z}}$, rate $B$ and $n$.
The codebook is revealed to the relay and the decoder but not to the transmitter. Once receiving $y^n$ the relay finds an index $w$ such that $z^n(w)$ is jointly typical with $y^n$. 
Such $w$ exists if
     $B > I(\rv{Y};\rv{Z}) + \delta(\epsilon')$,
where $\delta(\epsilon') $ tends to zero as $\epsilon' \rightarrow 0$.
\paragraph{Mismatched Decoding}
Once having $w$, the receiver finds a unique message $\hat{m}$ such that
\begin{equation}
    \hat{m} = \argmax_{m\in \set{M}} V^n(z^n(w)|x^n(m)).
\end{equation}
If there is none or more than one such message -- it declares an error event $\event{E}$. 

\paragraph{Analysis of the probability of error at the mismatched decoder}

Assume \acrshort{wlog} that $\rv{M} = 1$ was sent, and $\rv{W} = g(\rv{Y}^n )$ was chosen at the relay. The decoder makes an error if $V^n(\rv{Z}^n(\rv{W})|\rv{X}^n(m)) > V^n(\rv{Z}^n(\rv{W})|\rv{X}^n(1))  $ for some $m \neq 1$. Thus, the upper bound on the probability of error will be as follows
\begin{align}
    P_e
    &= \Prob{ \bigcup_{m=2}^{2^{nR}} V^n(\rv{Z}^n(\rv{W})|\rv{X}^n(m)) \geq V^n(\rv{Z}^n(\rv{W})|\rv{X}^n(1))  } \\
    &\leq (2^{nR}-1) \Prob{  V^n(\rv{Z}^n(\rv{W})|\rv{X}^n(2)) \geq V^n(\rv{Z}^n(\rv{W})|\rv{X}^n(1))  }.
\end{align}
We denote $\rv{X}^n \triangleq \rv{X}^n(1)$ and $\bar{\rv{X}}^n \triangleq \rv{X}^n(2)$. Thus,
\begin{equation}
    P_e \leq 2^{nR} \cdot \Prob{  V^n(\rv{Z}^n(\rv{W})|\bar{\rv{X}}^n) \geq V^n(\rv{Z}^n(\rv{W})|\rv{X}^n)  }.
\end{equation}
Further note that
    $\log V^n(z^n|x^n)
    = n \Exp[\epmf{\bv{x}\bv{z}}]{\log V(\rv{Z}|\rv{X})} $.
Denote $\pmf{} \triangleq \pmf{\rv{X}^n,\rv{Z}^n}$ and $\pmfT{} \triangleq \pmf{\bar{\rv{X}}^n,\rv{Z}^n}$. Note that by construction $\pmfT{\rv{X}} = \pmf{\rv{X}} $ and $\pmfT{\rv{Z}} = \pmf{\rv{Z}} $ .  It follows that
\begin{align}
    &\Prob{  V^n(z^n(W)|\bar{\rv{X}}^n) \geq V^n(z^n(W)|x^n)  } \\
    &= \Prob{ \Exp[\pmfT{}]{\log V(\rv{Z}|\rv{X})} > \Exp[\pmf{}]{\log V(\rv{Z}|\rv{X})}} \\
    &= \sum_{
    \substack{
    \pmfT{} \in \set{P}_{\set{X}\times \set{Z}} \colon \pmfT{\rv{X}} = \pmf{\rv{X}}, \pmfT{\rv{Z}} = \pmf{\rv{Z}}  \\
    \Exp[\pmfT{}]{\log V(\rv{Z}|\rv{X})} > \Exp[\pmf{}]{\log V(\rv{Z}|\rv{X})}
    }
    } \Prob{ (\bar{\rv{X}}^n,z^n) \in \typeclass{n}{\pmfT{}} }  \\
    &\eqann{\leq}{a} \sum_{
    \substack{
    \pmfT{} \in \set{P}_{\set{X}\times \set{Z}} \colon \pmfT{\rv{X}} = \pmf{\rv{X}}, \pmfT{\rv{Z}} = \pmf{\rv{Z}} \\
    \Exp[\pmfT{}]{\log V(\rv{Z}|\rv{X})} > \Exp[\pmf{}]{\log V(\rv{Z}|\rv{X})}
    }
    } 
    e^{-n \kld{\pmfT{} }{\pmf{\rv{X}} \times \pmf{\rv{Z}}}} \\
    &\leq (n+1)^{|\set{X}|\cdot |\set{Z}|} \max_{
    \substack{
    \pmfT{} \in \set{P}_{\set{X}\times \set{Z}} \colon \pmfT{\rv{X}} = \pmf{\rv{X}}, \pmfT{\rv{Z}} = \pmf{\rv{Z}} \\
    \Exp[\pmfT{}]{\log V(\rv{Z}|\rv{X})} > \Exp[\pmf{}]{\log V(\rv{Z}|\rv{X})}
    }
    } 
    e^{-n \kld{\pmfT{} }{\pmf{\rv{X}} \times \pmf{\rv{Z}}}} \\
    &= (n+1)^{|\set{X}|\cdot |\set{Z}|} 
    e^{-n J_{V} (\rv{X};\rv{Z})}.
\end{align}
where \eqannref{a} follows since $\tilde{\rv{X}}$ is independent of $z^n$.

The question is what is $\pmf{}$. The following section shows that with high probability we can take $\pmf{}$ as $\pmf{\rv{X}\rv{Z}}$,
\subsection{Joint Typicality of the Markov Tripple}
Since adding constraints reduces probability, then
\begin{multline}
    \Prob{(\rv{X}^n(1), \rv{Z}^n(\rv{W}(\rv{Y}^n))) \in \typset{\epsilon}{n}(\pmf{\rv{X}\rv{Z}})}
    \\
    \geq \Prob{(\rv{X}^n(1), \rv{Y}^n, \rv{Z}^n(\rv{W}(\rv{Y}^n))) \in \typset{\epsilon}{n}(\pmf{\rv{X} \rv{Y} \rv{Z}})}.
\end{multline}
The encoder at the relay uses joint typicality encoding, therefore
\begin{equation}
    \lim_{n \rightarrow \infty} \Prob{(y^n,\rv{Z}^n) \in \typset{\epsilon'}{n} (\pmf{\rv{Y}\rv{Z}}) } = 1.
\end{equation}
Furthermore, for every $z^n \in \typset{\epsilon'}{n}(\rv{Z}|y^n) $
\begin{align}
    &\Prob{\rv{Z}^n(\rv{W}) = z^n | \rv{Y}^n=y^n} \\
    &= \Prob{\rv{Z}^n(\rv{W}) = z^n, \rv{Z}^n(\rv{W}) \in \typset{\epsilon'}{n}(\rv{Z}|y^n)  | \rv{Y}^n=y^n} \\
    &\leq \Prob{\rv{Z}^n(\rv{W}) = z^n  | \rv{Y}^n=y^n, \rv{Z}^n(\rv{W}) \in \typset{\epsilon'}{n}(\rv{Z}|y^n)} \\
    &= 2^{-n(H(\rv{Z}|\rv{Y})-\delta(\epsilon'))}.
\end{align}
Similarly
\begin{equation}
    \Prob{\rv{Z}^n(\rv{W}) = z^n | \rv{Y}^n=y^n} \geq (1-\epsilon') 2^{-n(H(\rv{Z}|\rv{Y})-\delta(\epsilon'))}.
\end{equation}
Thus, the conditions of the Markov Lemma \cite[Lemma 12.1]{gamal2011} are satisfied and therefore
\begin{equation}
    \lim_{n \rightarrow \infty} \Prob{(\rv{X}^n(1), \rv{Y}^n, \rv{Z}^n(\rv{W}(\rv{Y}^n))) \in \typset{\epsilon}{n}} = 1.
\end{equation}

%% file: sections/mmib/converse.tex
Assume that the rate $R$ is larger than the random coding capacity. We then show that the random coding error probability tends to 1. Namely, the presumption is that the achievable rate $R$ is such that it is larger than all $J_V(\rv{X};\rv{Z})$ whenever $I(\rv{Y};\rv{Z})\leq B$. The additional constraint of $I(\rv{Y};\rv{Z})\leq B$ is the complication compared to the regular case.

We use random coding both at the transmitter and the relay.
Fix $\epsilon > 0$, and let $\typset{\epsilon}{n}(\pmf{\rv{X} \rv{Y}})$ and $\typset{\epsilon}{n}(\pmf{\rv{Z}})$ denote the $\epsilon$--typical sets associated with the input and output marginals, respectively. Similarly, fix $\delta > 0$, and let 
\begin{equation}
    \set{S}_{\delta ^-}^{(n)} \triangleq \left\{ (x^n,z^n) \colon \log V(z^n|x^n) \leq n(-D+\delta) \right\}.
\end{equation}
Assume without loss of generality that message $\rv{M}=1$ was sent and the index $\rv{W}$ was chosen at the relay. The average probability of correct message decoding can be upper bounded as follows:
\begin{align}
    &1 - \Bar{P_e} \\
    &= \Prob{V(\rv{Z}^n|\rv{X}^n(m)) < V(\rv{Z}^n|\rv{X}^n(1)) \forall m \neq 1 } \\
    &\leq  \Prob{
    \substack{
    V(\rv{Z}^n|\rv{X}^n(m)) < V(\rv{Z}^n|\rv{X}^n(1)) \forall m \in [2:2^{nR}], \\ (\rv{X}^n(1),\rv{Z}^n) \in \set{S}_{\delta ^-}^{(n)}, \rv{Y}^n \in \typset{\epsilon}{n}(\pmf{\rv{Y}}) , \rv{Z}^n \in \typset{\epsilon}{n}(\pmf{\rv{Z}})} } \nonumber \\
    &\phantom{=} + \Prob{\rv{Y}^n \notin \typset{\epsilon}{n}(\pmf{\rv{Y}})}
    + \Prob{\rv{Z}^n \notin \typset{\epsilon}{n}(\pmf{\rv{Z}})} \nonumber \\
    &\phantom{=} + \Prob{(\rv{X}^n(1),\rv{Z}^n) \notin \set{S}_{\delta ^-}^{(n)}}.
\end{align}
Since $\pmf{\rv{Y}^n}(y^n) = \prod_{i=1}^n \pmf{\rv{Y}}(y_i)$ and $\pmf{\rv{Z}^n}(z^n) = \prod_{i=1}^n \pmf{\rv{Z}}(z_i)$, then by the \acrshort{lln}, $\Prob{\rv{Y}^n \notin \typset{\epsilon}{n}(\pmf{\rv{Y}}) }$ and $\Prob{\rv{Z}^n \notin \typset{\epsilon}{n}(\pmf{\rv{Z}}) }$ tend to zero as $n\rightarrow \infty$. 

Consider the complement of the last term.
\begin{align}
    &\Prob{(\rv{X}^n(1),\rv{Z}^n) \in \set{S}_{\delta ^-}^{(n)}} \\
    &\geq 
    \Prob{(\rv{X}^n(1),\rv{Z}^n) \in \set{S}_{\delta ^-}^{(n)}, \left(\rv{X}^n(1),\rv{Y}^n,\rv{Z}^n(\rv{W}) \right) \in \typset{\epsilon}{n} } \\
    &=
    \Prob{ \left(\rv{X}^n(1),\rv{Y}^n,\rv{Z}^n(\rv{W}) \right) \in \typset{\epsilon}{n}(\pmf{\rv{X}\rv{Y}\rv{Z}}) } \\
    & \times \nonumber
    \Prob{(\rv{X}^n(1),\rv{Z}^n) \in \set{S}_{\delta ^-}^{(n)} \bigg| \left(\rv{X}^n(1) ,\rv{Y}^n,\rv{Z}^n(\rv{W}) \right) \in \typset{\epsilon}{n} }.
\end{align}
Consider the complement of the first multiplicand,
\begin{align}
    &\Prob{\left( \rv{X}^n(1),\rv{Y}^n, \rv{Z}^n(\rv{W}) \right) \notin \typset{\epsilon}{n}(\pmf{\rv{X}\rv{Y}\rv{Z}})} \\
    &= \Prob{
    \substack{
    \left\{ \left( \rv{X}^n(1),\rv{Y}^n \right) \notin \typset{\epsilon}{n} \right\} \\
    \bigcup \left\{ \left\{ \left( \rv{X}^n(1),\rv{Y}^n, \rv{Z}^n(\rv{W}) \right) \notin \typset{\epsilon}{n} \right\} \cap \left\{ \left( \rv{X}^n(1),\rv{Y}^n \right) \in \typset{\epsilon}{n} \right\} \right\} 
    }} \\
    &\leq \Prob{ \left\{ \left( \rv{X}^n(1),\rv{Y}^n \right) \notin \typset{\epsilon}{n} \right\} } \nonumber \\
    &+ \Prob{
    \substack{ \left\{ \left( \rv{X}^n(1),\rv{Y}^n, \rv{Z}^n(\rv{W}) \right) \notin \typset{\epsilon}{n} \right\} \\
    \cap \left\{ \left( \rv{X}^n(1),\rv{Y}^n \right) \in \typset{\epsilon}{n} \right\}  
    }} .
\end{align}
Since $\pmf{\rv{X}^n(1)\rv{Y}^n}(x^n,y^n) = \prod_{i=1}^n \pmf{\rv{X}\rv{Y}} (x_i,y_i)$ then by the \acrshort{lln} the first term tends to zero as $n\rightarrow \infty$. Furthermore, since $\left( \rv{X}^n(1),\rv{Y}^n \right) \in \typset{\epsilon}{n}(\pmf{\rv{X}\rv{Y}})$ and $\rv{Z}^n (\rv{W})\sim \pmf{\rv{Z}^n|\rv{Y}^n}$, by Markov Lemma \cite[Lemma 12.1]{gamal2011}, the second term also goes to zero as $n\rightarrow \infty$.
Next consider the second multiplicand, and take $\epsilon$ small enough,
\begin{align}
    \Prob{(\rv{X}^n(1),\rv{Z}^n) \in \set{S}_{\delta ^-}^{(n)} \bigg| \left(\rv{X}^n(1) ,\rv{Y}^n,\rv{Z}^n(\rv{W}) \right) \in \typset{\epsilon}{n} } =1, 
\end{align}
where the last equality follows since $(x^n,z^n)$ are jointly typical.
So it remains to upper bound the first term, henceforth denoted by $ \Prob{\event{A}}$, by a vanishing quantity as well. Note that
\begin{align}
    &\Prob{\event{A}} \\
    &\leq \Prob{\log V(\rv{Z}^n|\rv{X}^n(m)) \leq n(-D+\delta) \forall  m \neq 1, \rv{Z}^n\in \typset{\epsilon}{n} } \\
    &= \mkern-10mu \sum_{z^n \in \typset{\epsilon}{n}} \mkern-20mu \pmf{\rv{Z}^n}(z^n)  \Prob{\log V(z^n|\rv{X}^n(m)) \leq n(-D+\delta) \forall m \neq 1} \\
    &= \sum_{z^n \in \typset{\epsilon}{n}} \pmf{\rv{Z}^n}(z^n) \left[1-  \mkern-20mu \sum_{x^n \colon \log V(z^n|x^n) > n(-D+\delta)} \mkern-40mu  \pmf{\rv{X}^n}(x^n) \right] ^{2^{nR}-1}.
\end{align}
Denoting $h(z^n) \triangleq \sum_{x^n \colon \log V(z^n|x^n) > n(-D+\delta)}  \pmf{\rv{X}^n}(x^n) $, and using the fact that $1-\alpha \leq e^{-\alpha}$ we obtain
\begin{align}
    \Prob{\event{A}}
    &\leq \sum_{z^n \in \typset{\epsilon}{n}(\pmf{\rv{Z}})} \pmf{\rv{Z}^n}(z^n) \exp \left( - h(z^n) \cdot 2^{nR} \right) \\
    &\leq \exp \left( - \min_{z^n \in \typset{\epsilon}{n}(\pmf{\rv{Z}})} h(z^n) \cdot 2^{nR} \right).
\end{align}
To complete the proof, we need to show that for every $z^n \in \typset{\epsilon}{n}(\pmf{\rv{Z}})$, $h(z^n)$ is exponentially no smaller than $2^{-n I_V(\rv{X};\rv{Z})}$ when $\epsilon$ vanishes, and hence for every $R > I_V(\rv{X};\rv{Z})$, $\Prob{\event{A}}$ is essentially less than $\exp\left(-e^{n(R-I_V(\rv{X};\rv{Z}))} \right) \rightarrow 0$. To this end, let us further lower bound $h(z^n)$. First, note that for every $z^n \in \typset{\epsilon}{n}(\pmf{\rv{Z}})$,
\begin{align}
    h(z^n) 
    &= \sum_{x^n \colon \log V(z^n|x^n) > n(-D+\delta)}  \pmf{\rv{X}^n}(x^n) \\
    &= \sum_{ \set{T}(\epmf{\bv{x}|\bv{z}}) \colon \log V(\bv{z}|\bv{x}) \geq n(-D+\delta)} \mkern-60mu |\set{T}(\epmf{\bv{x}|\bv{z}})| \cdot 2^{-n[\kld{\epmf{\bv{x}}}{\pmf{\rv{X}}}+H(e\pmf{\bv{x}})]} \\
    &\geq \max_{ \set{T}(\epmf{\bv{x}|\bv{z}}) \colon \log V(\bv{z}|\bv{x}) \geq n(-D+\delta)} \mkern-60mu |\set{T}(\epmf{\bv{x}|\bv{z}})| \cdot 2^{-n[\kld{\pmf{\bv{x}}}{\pmf{\rv{X}}}+H(\epmf{\bv{x}})]} .
\end{align}
Note that since $z^n$ is given, $\set{T}(\epmf{\bv{x}|\bv{z}
})$ defines both $\epmf{\bv{x}}$ and $\epmf{\bv{x}|\bv{z}}$. Now, let
\begin{equation}
    \set{S} \triangleq \left\{ x^n \colon \log V(z^n|x^n) > n(-D+\delta) \right\} \cap \typset{\epsilon}{n}(\pmf{\rv{X}}).
\end{equation}
Therefore
\begin{align}
    h(z^n) 
    &\geq \max_{\set{T}(\epmf{\bv{x}|\bv{z}}) \in \set{S}} |\set{T}(\epmf{\bv{x}|\bv{z}})| \cdot 2^{-n[\kld{\pmf{\bv{x}}}{\pmf{\rv{X}}}+H(\epmf{\bv{x}})]}  \\
    &\geq \max_{\set{T}(\epmf{\bv{x}|\bv{z}})\in \set{S}} 2^{n[H(\epmf{\bv{x}|\bv{z}})-\delta(\epsilon)]} \cdot 2^{-n[H(\rv{X})+\epsilon']} \\
    &= 2^{-n [\min_{\pmfQ{\rv{X}|\rv{Z}} \in \set{S} } I(\pmfQ{\rv{X}}, \pmf{\rv{X} |\rv{Z}} ) + \epsilon' + \delta(\epsilon)]}.
\end{align}
Note that the last bound is by construction true for every $z^n \in \typset{\epsilon}{n}(\pmf{\rv{Z}})$, therefore, the marginal distribution of $\rv{Z}$ must be in the vicinity of $\pmf{\rv{Z}} (z) = \sum_{(x,y) \in \set{X} \times \set{Y}} \pmf{\rv{Z}|\rv{Y}} (z|y) \pmfQ{\rv{Y}|\rv{X}} (y|x) = \sum_{(x,y) \in \set{X} \times \set{Y}} \pmf{\rv{Z}|\rv{Y}} (z|y) \pmf{\rv{Y}|\rv{X}} (y|x)$. Thus, the following equality is established:
\begin{equation}
    \min_{\pmfQ{\rv{X}|\rv{Z}} \in \set{S} \cap \typset{\epsilon}{n}(\pmf{\rv{Z}})}  I(\pmf{\rv{X}},\pmfQ{\rv{X}|\rv{Z}}) = J_V(\rv{X};\rv{Z}).
\end{equation}
This completes the proof of the converse.

%% file: sections/mmib/gmi_lower_bound_proof.tex
\subsection{Primal Form}
The event
   $ \left\{ V^n(x^n(m),z^n) > V^n(x^n(1),z^n) \right\}$
is equivalent to the event
\begin{equation}
    \left\{ 
        \Exp[\epmf{\bv{x}(m) \bv{z}}]{\log V(\rv{X},\rv{Z})} > \Exp[\epmf{\bv{x}(1) \bv{z}}]{\log V(\rv{X},\rv{Z})}
    \right\}.
\end{equation}  
We proceed to bound the probability of error at the decoder. Denote the joint empirical distribution of some sequence pairs $(x^n(1),z^n)$ as $\pmf{}$, and $(x^n(2),z^n)$ as $\pmfT{}$.
\begin{align}
    &\Prob{\event{E}_4 \cap \event{E}_2^c} \\
    &= \Prob{ \bigcup_{m=2}^{2^{nR}} V^n(\rv{X}^n(m),\rv{Z}^n) > V^n(\rv{X}^n(1),\rv{Z}^n) } \\
    &\leq 2^{nR} \cdot \Prob{  V^n(\rv{X}^n(2),\rv{Z}^n) > V^n(\rv{X}^n(1),\rv{Z}^n) } \\
    &= \Prob{ \Exp[\pmfT{}]{\log V(\rv{X},\rv{Z})} > \Exp[\pmf{}]{\log V(\rv{X},\rv{Z})} } \\
    &= \sum_{ 
    \substack{
    \tilde{\pmf{}} \in \set{P}_n (\set{X} \times \set{Z}) \\
    \Exp[\tilde{\pmf{}}]{\log V(\rv{X},\rv{Z})} > \Exp[\pmf{}]{\log V(\rv{X},\rv{Z})} }}  \mkern-40mu \Prob{ (\rv{X}^n(2) ,\rv{Z}^n) \in \typeclass{n}{\Tilde{\pmf{}}}}  \\
    &\leq \sum_{ 
    \substack{
    \tilde{\pmf{}} \in \set{P}_n (\set{X} \times \set{Z}) \\
    \Exp[\tilde{\pmf{}}]{\log V(\rv{X},\rv{Z})} > \Exp[\pmf{}]{ \log V(\rv{X},\rv{Z})} }} e^{-n \kld{\tilde{\pmf{}}}{\pmfQ{\rv{X}} \times \pmf{\rv{Z}}}   } \\
    &\leq (n+1)^{|\set{X}| \cdot |\set{Z}|}  \max_{ 
    \substack{
    \tilde{\pmf{}} \in \set{P}_n (\set{X} \times \set{Z}) \\
    \Exp[\tilde{\pmf{}}]{\log V(\rv{X},\rv{Z})} > \Exp[\pmf{}]{\log V(\rv{X},\rv{Z})} }} e^{-n \kld{\tilde{\pmf{}}}{\pmfQ{\rv{X}} \times \pmf{\rv{Z}}}   } .
\end{align}
Thus $\Prob{\event{E}_4 \cap \event{E}_2^c}$ tends to zero when $n$ tends to infinity if 
\begin{equation} \label{eq:mmdec_gmi_rate_proof1}
    R < \min_{ 
    \substack{
    \tilde{\pmf{}} \in \set{P}_n (\set{X} \times \set{Z}) \\
    \Exp[\tilde{\pmf{}}]{\log V(\rv{X},\rv{Z})} > \Exp[\pmf{}]{\log V(\rv{X},\rv{Z})} }} \kld{\tilde{\pmf{}}}{\pmfQ{\rv{X}} \times \pmf{\rv{Z}}} .
\end{equation}
This completes the proof of the primal part of \Cref{theorem:mmib_mismatched_decoder_gmi_rate}.
\subsection{Dual Form}
Consider the minimization problem from \eqref{eq:mmdec_gmi_rate_proof1}. Since the objective function is convex (it is a divergence), and the constraints are linear, then this is a convex optimization problem. The latter further implies that strong duality holds \cite{boyd2004}. The respective Lagrangian is given by
\begin{align}
    L(\pmfT{}, \lambda) 
    &= \kld{\pmfT{}}{\pmf{\rv{X}} \cdot \pmf{\rv{Z}}} 
    + \lambda  \Exp[\pmf{}] { \log V(\rv{X},\rv{Z})} \\
    &\phantom{=} - \lambda \Exp[\pmfT{}] { \log V(\rv{X},\rv{Z})} \nonumber \\
    &\phantom{=} + \sum_{z \in \set{Z}} \mu(z) \sum_{x \in \set{X}} \left[ \pmfT{}(x,z) - \pmf{\rv{Z}}(z) \right] \nonumber \\
    &= \Exp[\pmfT{}] { \log \frac{\pmfT{}}{\pmf{\rv{X}} \cdot \pmf{\rv{Z}}} }
    + \lambda  \Exp[\pmf{}] {\log V(\rv{X},\rv{Z})} \\
    &\phantom{=}  - \lambda \Exp[\pmfT{}] {\log V(\rv{X},\rv{Z})} \nonumber \\
    &+ \sum_{z \in \set{Z}} \mu(z) \left[\sum_{x \in \set{X}}  \pmfT{}(x,z) - \pmf{\rv{Z}}(z) \right] \nonumber.
\end{align}
The dual objective function is given by
\begin{equation}
    q(\lambda) = \min_{\pmfT{} \in \set{P}_n(\set{X}\times \set{Z})} L(\pmfT{}, \lambda).
\end{equation}
We find the minimum by determining the stationary point
\begin{align}
    \frac{\partial L}{\partial \pmfT{}(x,z)}
    &= \log \frac{\pmfT{}(x,z)}{\pmf{\rv{X}}(x) \cdot \pmf{\rv{Z}}(z)}  + 1 - \lambda \log V(x,z) + \mu(z) = 0.
\end{align}
Thus,
\begin{equation}
    \pmfT{}^* (x,z) = \frac{\pmf{\rv{X}}(x) \pmf{\rv{Z}}(z)  V(x,z)^\lambda }{\sum_{x' \in \set{X}} \pmf{\rv{X}}(x')   V(x',z)^\lambda}.
\end{equation}
Plugging $\pmfT{}^*$ in the Lagrangian we obtain the following dual objective function
\begin{equation}
    q(\lambda) = \sum_{(x,z) \in \set{X}\times \set{Z}} \pmf{}(x,z) \log \frac{V(x,z)^{\lambda}}{\sum_{x' \in \set{X}} \pmf{\rv{X}}(x') V(x',z)^{\lambda}}.
\end{equation}
Finally, due to strong duality, we have the following equality
\begin{equation}
    I_{\mathsf{GMI}}(\pmf{\rv{X} \rv{Z}} ) = \max_{\lambda \geq 0}  \sum_{(x,z) }  \pmf{}(x,z) \log \frac{V(x,z)^{\lambda}}{\sum_{x' \in \set{X}} \pmf{\rv{X}}(x') V(x',z)^{\lambda}},
\end{equation}
where $I_{\mathsf{GMI}}(\pmf{\rv{X} \rv{Z}} )$ is defined in \eqref{eq:mmibdec_Igmi_optimization_problem}.

%% file: sections/mmrib/lower_bound.tex
We use random coding. Let $\epsilon' < \epsilon$. Fix the marginal \acrshort{pmf}s $\pmf{\rv{X}}$ and $\pmf{\rv{Z}}$. This induces a marginal \acrshort{pmf} for $\rv{Y}$, i.e., $\pmf{\rv{Y}}(y) = \sum_{x \in \set{X}} \pmf{\rv{Y}|\rv{X}}(y|x)  \pmf{\rv{X}} (x)$.

\paragraph{Transmitter's Codebook Generation and Encoding}
Generate a random codebook $ \set{C} = \{ x^n(m) \} $, $m \in [1:2^{nR}]$ according to 
$\pmf{\rv{X}}$, rate $R$ and $n$.
The codebook is revealed to the encoder and the decoder but not the relay.

\paragraph{Mismatched Relay's Codebook Generation}
Generate a random codebook $ \set{C}_R = \{ z^n(w) \} $, $w \in [1:2^{nB}]$ according to 
$\pmf{\rv{Z}}$, rate $B$ and $n$.
The codebook is revealed to the relay and the decoder but not to the transmitter.

\paragraph{Mismatched Relay Encoding}
Upon observing $y^n$, the relay finds an index $\hat{w} \in [1:2^{nB}]$  such that 
\begin{equation}
    \hat{w} = \argmin_{w \in \set{W}} d_0^n(y^n,z^n(w)).
\end{equation}
The relay conveys the selected index $\hat{w}$ to the decoder.

\paragraph{Decoding}
Note that the decoder does not know the true channel realization from $\rv{X}$ to $\rv{Z}$. Since the encoder at the relay applies a fixed decoding metric instead of joint-typicality encoding, it does not give rise to the inherent ``test-channel" as usually the case in typicality-based schemes. This situation resembles the setting of universal decoder, where the decoder only knows the codebook. With this observation in mind, we suggest the following decoder. Once acquiring $\hat{w}$ from the relay, the receiver finds a unique message $\hat{m}$ such that
\begin{equation}
    \hat{m} = \argmax_{m\in \set{M}} I(\epmf{\bv{x}(m) \bv{z}(\hat{w})}),
\end{equation}
where $I(\epmf{\bv{x}(m) \bv{z}(\hat{w})})$ denotes the mutual information evaluated for the joint empirical distribution of the $m$th codeword $x^n(m)$ and $z^n(\hat{w}))$.
If $\hat{m} \neq m$ an error event $\event{E}$ is declared.

\paragraph{Analysis of the performance at the mismatched relay}
The relay observes $y^n$. Since $\pmf{\rv{Y}^n}(y^n) = \prod_{i=1}^n \pmf{\rv{Y}}(y_i)$, by the \acrshort{lln}, with high probability $y^n \in \typset{\epsilon}{n}(\pmf{\rv{Y}})$. 
We define $\pmf{\rv{Z},n} \in \set{P}_n(\set{Z})$ to be an arbitrary type having the same support as $\pmf{\rv{Z}}$, and satisfying:
  $  \lVert \pmf{\rv{Z}} - \pmf{\rv{Z},n} \rVert_{\infty} \leq \frac{1}{n}$.
Note that for every $w \in \set{W}$, $z^n(w) \in \typset{\epsilon}{n}(\pmf{\rv{Z}})$, such that $\epmf{z^n(w)} = \pmf{\rv{Z},n}$. Fix $y^n$ and $w$ and consider the distortion metric:
\begin{equation}
    d_0(y^n,z^n(w))
    = \frac{1}{n} \sum_{i=1}^n d_0(y_i,z_i) 
    = \Exp[\epmf{\bv{y} \bv{z}(w)}]{d_0(\rv{Y},\rv{Z})}.
\end{equation}
We will show next that $\epmf{\bv{y} \bv{z}}$ has to satisfy some specific structure.

The rest of the analysis here follows the one from \cite[Sec. 4.5]{Scarlett2020}. Define the following sets:
\begin{equation*}
    \set{S}_n^{(\minus)} \mkern-5mu = \mkern-5mu \left\{ \mkern-5mu \pmfQ{\rv{Y}\rv{Z}} \mkern-5mu \in \mkern-5mu \set{P}_{\set{Y}\times \set{Z}}^n \colon \mkern-5mu
    \pmfQ{\rv{Y}} \mkern-5mu =  \mkern-5mu \pmf{\rv{Y}}, \mkern-5mu \pmfQ{\rv{Z}} \mkern-5mu =  \mkern-5mu \pmf{\rv{Z},n}, I(\pmfQ{\rv{Y}\rv{Z}}) \mkern-5mu \geq \mkern-5mu B \mkern-5mu + \mkern-5mu \delta 
    \right\} ,
\end{equation*}
and
\begin{equation*}
    \set{S}_n^{(+)} \mkern-5mu = \mkern-5mu \left\{ \mkern-5mu \pmfQ{\rv{Y}\rv{Z}} \mkern-5mu \in \mkern-5mu \set{P}_{\set{Y}\times \set{Z}}^n \colon \mkern-5mu
    \pmfQ{\rv{Y}} \mkern-5mu =  \mkern-5mu \pmf{\rv{Y}}, \pmfQ{\rv{Z}}  \mkern-5mu =\mkern-5mu \pmf{\rv{Z},n}, \mkern-5mu I(\pmfQ{\rv{Y}\rv{Z}}) \mkern-5mu \leq \mkern-5mu B- \delta 
    \right\},
\end{equation*}
where $\set{P}_{\set{Y} \times \set{Z}} ^n$ is the subset of joint \acrshort{pmf}s on $\set{Y} \times \set{Z}$ that corresponds to the joint type of some length-$n$ sequence $(y^n,z^n)$.
\begin{lemma} The union of joint types of the pairs $(y^n,z^n(w))$ satisfy
\begin{enumerate}
    \item 
      $  \lim_{n \rightarrow \infty} \Prob{ \bigcup_{w = 1}^{2^{nB}} \epmf{\bv{y} \bv{z}(w)} \in \set{S}_n^{(-)}} = 0$.
    \item 
        $\lim_{n \rightarrow \infty} \Prob{\bigcup_{w = 1}^{2^{nB}} \epmf{\bv{y} \bv{z}(w)} \in \set{S}_n^{(+)}} = 1$.
\end{enumerate}

\end{lemma}
\begin{IEEEproof}
Clearly,
    \begin{align}
        &\Prob{\bigcup_{w=1}^{2^{nB}} \left\{ (y^n,\rv{Z}^n(w)) \in \typeclass{n}{\pmfQ{\rv{Y}\rv{Z}}} \right\} } \nonumber \\
        &= 1-\prod_{w=1}^{2^{nB}} \Prob{ (y^n,\rv{Z}^n(w)) \notin \typeclass{n}{\pmfQ{\rv{Y}\rv{Z}}}  } \\
        &= 1-\left[ 1- \Prob{ (y^n,\rv{Z}^n) \in \typeclass{n}{\pmfQ{\rv{Y}\rv{Z}}}  } \right]^{2^{nB}} .
        \label{eq:mmrib_direct_proof_event1}
    \end{align}
    Note that $(1-x)^k \leq e^{-kx}$ for $x \in [0,1]$ and $k \geq 0$. Furthermore, 
    \begin{equation}
        \lim_{n \rightarrow \infty} \left( 1+ \frac{z}{n} \right)^{n} = e^z.
    \end{equation}
    Consider the following cases:
    \begin{itemize}
        \item If $\Prob{ (y^n,\rv{Z}^n) \in \typeclass{n}{\pmfQ{\rv{Y}\rv{Z}}}  } \geq 2^{-n(B-\epsilon)}$, then the \acrshort{rhs} of \eqref{eq:mmrib_direct_proof_event1} is lower bounded by
        \begin{multline}
            \Prob{\bigcup_{w=1}^{2^{nB}} \left\{ (y^n,\rv{Z}^n(w)) \in \typeclass{n}{\pmfQ{\rv{Y}\rv{Z}}} \right\} } \\
            \geq 1-[(1-2^{-n(B-\epsilon)})^{2^{n(B-\epsilon)}}]^{2^{n\epsilon}} .
        \end{multline}
        Taking the limit of $n$ tends to $\infty$ we have
        \begin{align}
            &\lim_{n \rightarrow \infty} \Prob{\bigcup_{w=1}^{2^{nB}} \left\{ (y^n,\rv{Z}^n(w)) \in \typeclass{n}{\pmfQ{\rv{Y}\rv{Z}}} \right\} } \nonumber \\
            &\geq \lim_{n \rightarrow \infty} 1-[(1-2^{-n(B-\epsilon)})^{2^{n(B-\epsilon)}}]^{2^{n\epsilon}} \\
            &= \lim_{n \rightarrow \infty}  1- e^{-2^{n\epsilon}} = 1.
        \end{align}
        \item If $\Prob{ (y^n,\rv{Z}^n) \in \typeclass{n}{\pmfQ{\rv{Y}\rv{Z}}}  } \leq 2^{-n(B+\epsilon)}$, then the \acrshort{rhs} of \eqref{eq:mmrib_direct_proof_event1} is upper bounded by
        \begin{multline}
            \Prob{\bigcup_{w=1}^{2^{nB}} \left\{ (y^n,\rv{Z}^n(w)) \in \typeclass{n}{\pmfQ{\rv{Y}\rv{Z}}} \right\} } \\
            \leq 1-[(1-2^{-n(B+\epsilon)})^{2^{n(B+\epsilon)}}]^{2^{-n\epsilon}} .
        \end{multline}
        Taking the limits of $n$ tends to $\infty$ we have
        \begin{align}
            &\lim_{n \rightarrow \infty} \Prob{\bigcup_{w=1}^{2^{nB}} \left\{ (y^n,\rv{Z}^n(w)) \in \typeclass{n}{\pmfQ{\rv{Y}\rv{Z}}} \right\} } \nonumber \\
            &\leq  \lim_{n \rightarrow \infty} 1-[(1-2^{-n(B+\epsilon)})^{2^{n(B+\epsilon)}}]^{2^{-n\epsilon}} \\
            &=  \lim_{n \rightarrow \infty}  1- e^{-2^{-n\epsilon}} 
            = \lim_{n \rightarrow \infty}  1- 1 + 2^{-n\epsilon} = 0.
        \end{align}
    \end{itemize}
    Further, note that since $\pmfQ{\rv{Y}} = \epmf{y^n}$ and $\pmfQ{\rv{Z}} = \pmf{\rv{Z},n}$, we have
    \begin{equation}
        2^{-n[I(\pmfQ{\rv{Y}\rv{Z}}) + \delta]} \leq \Prob{ (y^n,\rv{Z}^n) \in \typeclass{n}{\pmfQ{\rv{Y}\rv{Z}}}  } \leq 2^{-n[I(\pmfQ{\rv{Y}\rv{Z}}) - \delta]}.
    \end{equation}
    Thus, if $ I(\pmfQ{\rv{Y}\rv{Z}}) \leq B - \delta' $ then 
    \begin{equation}
        \lim_{n \rightarrow \infty} \Prob{\bigcup_{w = 1}^{2^{nB}} \epmf{\bv{y} \bv{z}(w)} \in \set{S}_n^{(+)}} = 1.
    \end{equation}
    Otherwise, if $ I(\pmfQ{\rv{Y}\rv{Z}}) \geq B + \delta' $ then 
    \begin{equation}
       \lim_{n \rightarrow \infty} \Prob{ \bigcup_{w = 1}^{2^{nB}} \epmf{\bv{y} \bv{z}(w)} \in \set{S}_n^{(-)}} = 0.
    \end{equation}
\end{IEEEproof}
Thus, $\epmf{\bv{y}\bv{z}} \in \set{S}_n^{(+)}$ with high probability.
\paragraph{Analysis of the probability of error at the decoder}
Note that due to symmetry of codebook generation
\begin{align}
    &\Prob{\hat{\rv{M}} \neq \rv{M}}  \nonumber
    =  \Prob{\hat{\rv{M}} \neq \rv{M} | \rv{M}=1} \\
    &\leq \Prob{I(\epmf{\rv{X}^n(1) \rv{Z}^n(W)}) < \theta |\rv{M}=1} \\
    &+ \Prob{I(\epmf{\rv{X}^n(m) \rv{Z}^n(W)}) \geq \theta  \text{ for some } m \neq 1 |\rv{M}=1} \nonumber \\
    &\leq \Prob{I(\epmf{\rv{X}^n(1) \rv{Z}^n(W)}) < \theta |\rv{M}=1} \nonumber \\
    &+ 2^{nR} \cdot \Prob{I(\epmf{\rv{X}^n(2) \rv{Z}^n(W)}) \geq \theta  |\rv{M}=1}.
\end{align}
This implies that we can assume without loss of generality that $\rv{M} = 1$  is sent.
Consider the first term. Note that for every $x^n \in \typset{\epsilon''}{n}(\pmf{\rv{x}})$, $y^n \in \typset{\epsilon'}{n}(\pmf{\rv{Y}|\rv{X}}) $ and $z^n \in \typeclass{n}{\pmfQ{\rv{Z}|\rv{Y}}}$ we have
\begin{align}
    &\epmf{\bv{x} \bv{y} \bv{z}} (a,b,c) \\
    &= \frac{\sum_{i=1}^n \indicator{x_i=a,y_i=b,z_i=c}}{n} \\
    &= \frac{\sum_{i=1}^n \indicator{x_i=a}}{n} \frac{\sum_{i=1}^n \indicator{x_i=a,y_i=b,z_i=c}}{\sum_{i=1}^n \indicator{x_i=a}} \\
    &= \epmf{\bv{x}}(a) \epmf{\bv{y}|\bv{x}}(b|a) \pmfQ{\rv{Z}|\rv{Y}}(c|b) .
\end{align}
Thus,
    $\epmf{x^n (1)z^n(W)} = \epmf{x^n(1) y^n} \circ \pmfQ{\rv{Z}|\rv{Y}} $,
and
\begin{equation}
    \lim_{n \rightarrow \infty} I(\epmf{x^n (1)z^n}) = I(\pmf{\rv{X} \rv{Y}} \circ \pmfQ{\rv{Z}|\rv{Y }}).
\end{equation}
Thus,
\begin{equation}
    \lim_{n \rightarrow \infty} I(\epmf{\rv{X}^n (1)\rv{Z}^n(W)}) \geq \min_{\pmfQ{\rv{Z}|\rv{Y}} \in \set{S}_n^{(+)} } I(\pmf{\rv{X} \rv{Y}} \circ \pmfQ{\rv{Z}|\rv{Y }}) >\theta.
\end{equation}
In such case
    $\lim_{n \rightarrow \infty} \Prob{I(\pmf{\rv{X}^n(1) \rv{Z}^n(W)}) < \theta |\rv{M}=1}  = 0$.
Consider the second term. Let
\begin{equation}
    \set{A}_n \triangleq \left\{ \pmfQ{\rv{X}\rv{Z}} \in \set{P}_{\set{X} \times \set {Z}}^n \colon I(\pmfQ{\rv{X}\rv{Z}}) \geq \theta \right\}.
\end{equation}
Since for $m\neq 1$ we have $\pmf{\rv{X}^n(m) \rv{Z}^n(W)}(x^n,z^n) = \prod_{i=1}^n \pmf{\rv{X}}(x_i) \pmf{\rv{Z}} (z_i)$, then by Sanov's theorem \cite[Thm. 11.4.1]{Cover2006}
\begin{align}
    &\Prob{I(\pmf{\bv{X}(m) \bv{Z}(W)}) \geq \theta \big| \rv{M}=1}  \\
    &\leq (n+1)^{|\set{X}| \cdot |\set{Z}|}2^{-n \min_{\pmfQ{\rv{X}\rv{Z}} \in \set{A}_n} D(\pmfQ{\rv{X}\rv{Z}} \| \pmf{\rv{X}} \pmf{\rv{Z}})} \\
    &\leq (n+1)^{|\set{X}| \cdot |\set{Z}|}2^{-n \theta}.
\end{align}
Thus, the second term goes to zero as $ n \rightarrow \infty $ if $R < \theta$. This completes the proof of the theorem.

%% file: main.bbl
\begin{thebibliography}{10}

\bibitem{Sanderovich2008a}
A.~{Sanderovich}, S.~{Shamai}, Y.~{Steinberg}, and G.~{Kramer}, ``Communication
  via decentralized processing,'' {\em IEEE Trans. Inf. Theory}, vol.~54,
  pp.~3008--3023, July 2008.

\bibitem{EstellaAguerri2019}
I.~Estella~Aguerri, A.~Zaidi, G.~Caire, and S.~Shamai~Shitz, ``On the capacity
  of cloud radio access networks with oblivious relaying,'' {\em IEEE Trans.
  Inf. Theory}, vol.~65, pp.~4575--4596, July 2019.

\bibitem{Richardson2008}
T.~Richardson and R.~Urbanke, {\em Modern {C}oding {T}heory}.
\newblock Cambridge, U.K.: Cambridge Univ. Press, 2008.

\bibitem{Tishby1999}
N.~Tishby, F.~C.~N. Pereira, and W.~Bialek, ``The information bottleneck
  method,'' in {\em 37th Annu. Allerton Conf. Commun. Control Comput.},
  pp.~368–--377, Sept. 1999.

\bibitem{zaidi2020information}
A.~Zaidi, I.~Estella-Aguerri, and S.~Shamai~(Shitz), ``On the information
  bottleneck problems: Models, connections, applications and information
  theoretic views,'' {\em Entropy}, vol.~22, no.~2, 2020.

\bibitem{Goldfeld2020}
Z.~Goldfeld and Y.~Polyanskiy, ``The information bottleneck problem and its
  applications in machine learning,'' {\em IEEE Journal on Selected Areas in
  Information Theory}, vol.~1, pp.~19--38, May 2020.

\bibitem{Aguerri2021}
I.~E. Aguerri and A.~Zaidi, ``Distributed variational representation
  learning,'' {\em IEEE Trans. Pattern Anal. Mach. Intell.}, vol.~43,
  pp.~120--138, Jan. 2021.

\bibitem{shannon1959coding}
C.~E. Shannon {\em et~al.}, ``Coding theorems for a discrete source with a
  fidelity criterion,'' {\em IRE Nat. Conv. Rec}, vol.~4, no.~142-163, p.~1,
  1959.

\bibitem{Dobrushin1962}
R.~{Dobrushin} and B.~{Tsybakov}, ``Information transmission with additional
  noise,'' {\em IRE Transactions on Information Theory}, vol.~8, pp.~293--304,
  Sept. 1962.

\bibitem{Wolf1970}
J.~Wolf and J.~Ziv, ``Transmission of noisy information to a noisy receiver
  with minimum distortion,'' {\em IEEE Trans. Inf. Theory}, vol.~16,
  pp.~406--411, July 1970.

\bibitem{Courtade2014b}
T.~A. {Courtade} and T.~{Weissman}, ``{M}ultiterminal {S}ource {C}oding {U}nder
  {L}ogarithmic {L}oss,'' {\em IEEE Trans. Inf. Theory}, vol.~60, pp.~740--761,
  Jan. 2014.

\bibitem{Painsky2018}
A.~Painsky and G.~Wornell, ``On the universality of the logistic loss
  function,'' in {\em 2018 IEEE International Symposium on Information Theory
  (ISIT)}, pp.~936--940, 2018.

\bibitem{Scarlett2020}
J.~Scarlett, A.~G. i~Fàbregas, A.~Somekh-Baruch, and A.~Martinez,
  ``Information-theoretic foundations of mismatched decoding,'' {\em
  Foundations and Trends® in Communications and Information Theory}, vol.~17,
  no.~2–3, pp.~149--401, 2020.

\bibitem{hui1983fundamental}
J.~Y.~N. Hui, {\em Fundamental issues of multiple accessing}.
\newblock PhD thesis, Massachusetts Institute of Technology, 1983.

\bibitem{Csiszar1981}
I.~Csiszar and J.~Korner, ``Graph decomposition: A new key to coding
  theorems,'' {\em IEEE Trans. Inf. Theory}, vol.~27, pp.~5--12, Jan. 1981.

\bibitem{Csiszar1995}
I.~Csiszar and P.~Narayan, ``Channel capacity for a given decoding metric,''
  {\em IEEE Trans. Inf. Theory}, vol.~41, pp.~35--43, Jan. 1995.

\bibitem{kaplan1993information}
G.~Kaplan and S.~Shamai, ``Information rates and error exponents of compound
  channels with application to antipodal signaling in a fading environment,''
  {\em AEU. Archiv f{\"u}r Elektronik und {\"U}bertragungstechnik}, vol.~47,
  no.~4, pp.~228--239, 1993.

\bibitem{Merhav1994}
N.~Merhav, G.~Kaplan, A.~Lapidoth, and S.~Shamai~Shitz, ``On information rates
  for mismatched decoders,'' {\em IEEE Trans. Inf. Theory}, vol.~40,
  pp.~1953--1967, Nov. 1994.

\bibitem{lapidoth1997}
A.~Lapidoth, ``On the role of mismatch in rate distortion theory,'' {\em IEEE
  Trans. Inf. Theory}, vol.~43, pp.~38--47, Jan. 1997.

\bibitem{zhou2019}
L.~Zhou, V.~Y.~F. Tan, and M.~Motani, ``Refined asymptotics for rate-distortion
  using gaussian codebooks for arbitrary sources,'' {\em IEEE Trans. Inf.
  Theory}, vol.~65, pp.~3145--3159, May 2019.

\bibitem{bai2022achievable}
L.~Bai, Z.~Wu, and L.~Zhou, ``Achievable refined asymptotics for successive
  refinement using gaussian codebooks,'' {\em arXiv preprint arXiv:2208.03926},
  2022.

\bibitem{Kanabar2022}
M.~Kanabar and J.~Scarlett, ``Mismatched rate-distortion theory: Ensembles,
  bounds, and general alphabets,'' {\em arXiv preprint arXiv:2203.15193}, 2022.

\bibitem{Choi2015}
J.~Choi, D.~J. Love, D.~R. Brown, and M.~Boutin, ``Quantized distributed
  reception for mimo wireless systems using spatial multiplexing,'' {\em IEEE
  Trans. Signal Process.}, vol.~63, pp.~3537--3548, July 2015.

\bibitem{Brown2013}
D.~R. Brown, M.~Ni, U.~Madhow, and P.~Bidigare, ``Distributed reception with
  coarsely-quantized observation exchanges,'' in {\em 2013 47th Annual
  Conference on Information Sciences and Systems (CISS)}, pp.~1--6, 2013.

\bibitem{liang2014joint}
L.~Liang, S.~Bi, and R.~Zhang, ``Joint power control and fronthaul rate
  allocation for throughput maximization in ofdma-based cloud radio access
  network,'' {\em arXiv preprint arXiv:1407.3855}, 2014.

\bibitem{Wang2013}
R.~Wang, D.~Richard~Brown, M.~Ni, U.~Madhow, and P.~Bidigare, ``Outage
  probability analysis of distributed reception with hard decision exchanges,''
  in {\em 2013 Asilomar Conference on Signals, Systems and Computers},
  pp.~597--601, 2013.

\bibitem{Blahut1972}
R.~Blahut, ``Computation of channel capacity and rate-distortion functions,''
  {\em IEEE Trans. Inf. Theory}, vol.~18, pp.~460--473, July 1972.

\bibitem{Arimoto1972}
S.~Arimoto, ``An algorithm for computing the capacity of arbitrary discrete
  memoryless channels,'' {\em IEEE Trans. Inf. Theory}, vol.~18, pp.~14--20,
  Jan. 1972.

\bibitem{Gallager1968}
R.~G. Gallager, {\em Information theory and reliable communication / Robert G.
  Gallager.}
\newblock New York: Wiley, 1968.

\bibitem{Somekh2022}
A.~Somekh-Baruch, ``Upper bounds on the mismatched reliability function and
  capacity using a genie receiver,'' {\em arXiv preprint arXiv:2203.08524},
  2022.

\bibitem{Xu2021b}
H.~Xu, T.~Yang, G.~Caire, and S.~Shamai~(Shitz), ``Information bottleneck for a
  rayleigh fading mimo channel with an oblivious relay,'' {\em Information},
  vol.~12, no.~4, 2021.

\bibitem{Xu2021}
H.~Xu, T.~Yang, G.~Caire, and S.~Shamai~Shitz, ``Information bottleneck for an
  oblivious relay with channel state information: the vector case,'' in {\em
  2021 IEEE International Symposium on Information Theory (ISIT)},
  pp.~2483--2488, 2021.

\bibitem{gamal2011}
A.~El~Gamal and Y.-H. Kim, {\em Network information theory}.
\newblock Cambridge university press, 2011.

\bibitem{boyd2004}
S.~Boyd and L.~Vandenberghe, {\em Convex Optimization}.
\newblock Cambridge University Press, 2004.

\bibitem{Cover2006}
T.~M. Cover and J.~A. Thomas, {\em Elements of {I}nformation {T}heory}.
\newblock Hoboken, NJ, USA: Wiley, 2006.

\end{thebibliography}
